\documentclass[reprint,aps,prl,twocolumn]{revtex4-2}
\UseRawInputEncoding

\usepackage{amsmath}
\usepackage{float}
\usepackage{graphicx}
\usepackage[utf8]{inputenc}
\usepackage{multirow}

\usepackage[colorlinks,citecolor=blue]{hyperref}

\renewcommand{\thetable}{\arabic{table}}

\begin{document}

\title{Quantum Teleportation from Telecom Photons to Erbium-ion Ensembles} 

\author{Yu-Yang An}
\thanks{Equal contribution}
\author{Qian He}
\thanks{Equal contribution}
\author{Wenyi Xue}
\thanks{Equal contribution}
\author{Ming-Hao Jiang}
\author{Chengdong Yang}
\author{Yan-Qing Lu}
\author{Shining Zhu}
\author{Xiao-Song Ma}
\email{Xiaosong.Ma@nju.edu.cn}
\affiliation{National Laboratory of Solid-state Microstructures, School of Physics, College of Engineering and Applied Sciences, Collaborative Innovation Center of Advanced Microstructures, Jiangsu Physical Science Research Center, Nanjing University, Nanjing 210093,China}
\affiliation{Synergetic Innovation Center of Quantum Information and Quantum Physics, University of Science and Technology of China, Hefei, Anhui 230026, China}
\affiliation{Hefei National Laboratory, Hefei 230088, China}



\date{\today}

\begin{abstract}

	To realize a quantum internet, the distribution of quantum states via quantum teleportation with quantum memories is a key ingredient. Being compatible with existing fiber networks, entangled photons and quantum memories at telecom-wavelength are of central interest for such a scalable quantum network. Here, we demonstrate quantum teleportation from a telecom-wavelength photonic qubit to a solid-state quantum memory based on erbium-ion ensembles, which have a native optical transition at 1.5 $\mu$m telecom C-band. To accomplish this, we use chip-scale silicon nitride micro-resonators to generate entangled photons with narrow linewidth, compatible with the quantum memory. We confirm the quality of the quantum teleportation procedure using quantum state and process tomography techniques, in which both the quantum state and process fidelities exceeds the classical limit. These results pave the way for the realization of scalable quantum networks based on solid-state devices.
 
\end{abstract}

\maketitle

\section{Introduction}
	Quantum teleportation\cite{PhysRevLett.70.1895_1993,bouwmeester1997experimental,PhysRevLett.80.1121,science1998} allows the transfer of an unknown quantum state between distant physical systems. The combination of quantum teleportation with quantum memories provides scalable schemes of quantum networks\cite{kimble_quantum_2008,Hanson2018} for realizing distributed quantum computation and sensing. Light-to-matter quantum teleportation has been demonstrated using quantum memories based on trapped ions\cite{barrett_deterministic_2004,riebe_deterministic_2004}, atomic ensembles\cite{sherson_quantum_2006,krauter_deterministic_2013,chen_memory-built-quantum_2008,Xiao-HuiBao_pnas.2012}, remote single ions\cite{Olmschenk_science2009} and atoms\cite{Nolleke_PRL2013}, solid-state spin qubits\cite{gao_quantum_2013} and optomechanical systems\cite{fiaschi_optomechanical_2021}. One of the promising candidates for realizing quantum memory are rare-earth ions doped in solids, which have multimode storage capability\cite{AfzeliusPRA2009,Rakonjac_PRL2021,businger_non-classical_2022,wei_quantum_2024}, long coherence time\cite{zhong_optically_2015,ma_one-hour_2021} and high storage efficiency\cite{hedges_efficient_2010,Duranti_2024}. Previously, quantum teleportation has been realized with $\rm Nd^{3+}$\cite{bussieres_quantum_2014}, $\rm Pr^{3+}$\cite{lago-rivera_long_2023} and ${\rm Eu^{3+}}$\cite{liu_nonlocal_2024} ions.
    
    Among the various rare-earth ions, erbium is particularly important since it has optical transitions in the telecom C-band, which sits right at the low-loss transmission window for optical fibers and has been studied as a potential candidate for quantum memory\cite{Lauritzen_PRL2010}. Therefore, quantum teleportation would beneﬁt from compatibility with the existing telecommunication infrastructure. Note that single erbium ions have also been used as single-photon sources for emitting telecom single photons\cite{Dibos_PRL2018,Benjamin_PRX2020,Huang_CPL_2023,ourari_indistinguishable_2023,Yu_Yong_PRL2023,Gritsch2024optical_singles_hot_readout_spin} and recently for generating spin-photon entanglement\cite{uysal2024spinphotonentanglementsingleer3}. It has been recently shown that in a high magnetic field, isotopically purified ${\rm ^{167}Er^{3+}}$ ions with $I$ = 7/2 doped in YSO crystal possess a hyperfine coherence time of 1.3 s\cite{Rancic_coherence_2018}. This work rejuvenates the interest in using erbium ions for quantum memory\cite{Craiciu_PRApplied2019,Stuart_PRR_2021,Liu_PRL_2022,jiang_quantum_2023}.

    In this work, we demonstrate the quantum state teleportation from telecom photons to a solid-state quantum memory based on ${\rm ^{167}Er^{3+}}$ ions. Firstly, through the cavity enhanced spontaneous four-wave mixing (SFWM), we generate entangled photon pairs (signal and idler photons) from an integrated photonic chip based on a silicon nitride (SiN) microring resonator (MRR). Secondly, we send the idler photon to Alice, who performs the joint Bell-state measurement (BSM) on both idler and input photons\cite{Weinfurter_1994} obtained from attenuated laser. By doing so, we project both photons onto the Bell-state $|\Psi^{-}\rangle$. Thirdly, the signal photon is sent to the ${\rm ^{167}Er^{3+}}$ ions quantum memory, in which the quantum state of the signal photon is mapped onto that of the ensemble of ${\rm ^{167}Er^{3+}}$ ions and stored for more than 2 $\mu$s. After storage, we retrieve the quantum state of the signal photon from the Er ions and measure its quantum state.  Successful teleportation between telecom photons and Er ions is certiﬁed using quantum state and process tomography\cite{Daniel_PRA_2001,Nielsen_Chuang_2010}, respectively. See Note S1 of Ref.\cite{SI} for details on quantum teleportation protocol used in our work.

\section{Overview of the experimental setup}

\begin{figure*}
	\includegraphics[width=0.8\linewidth]{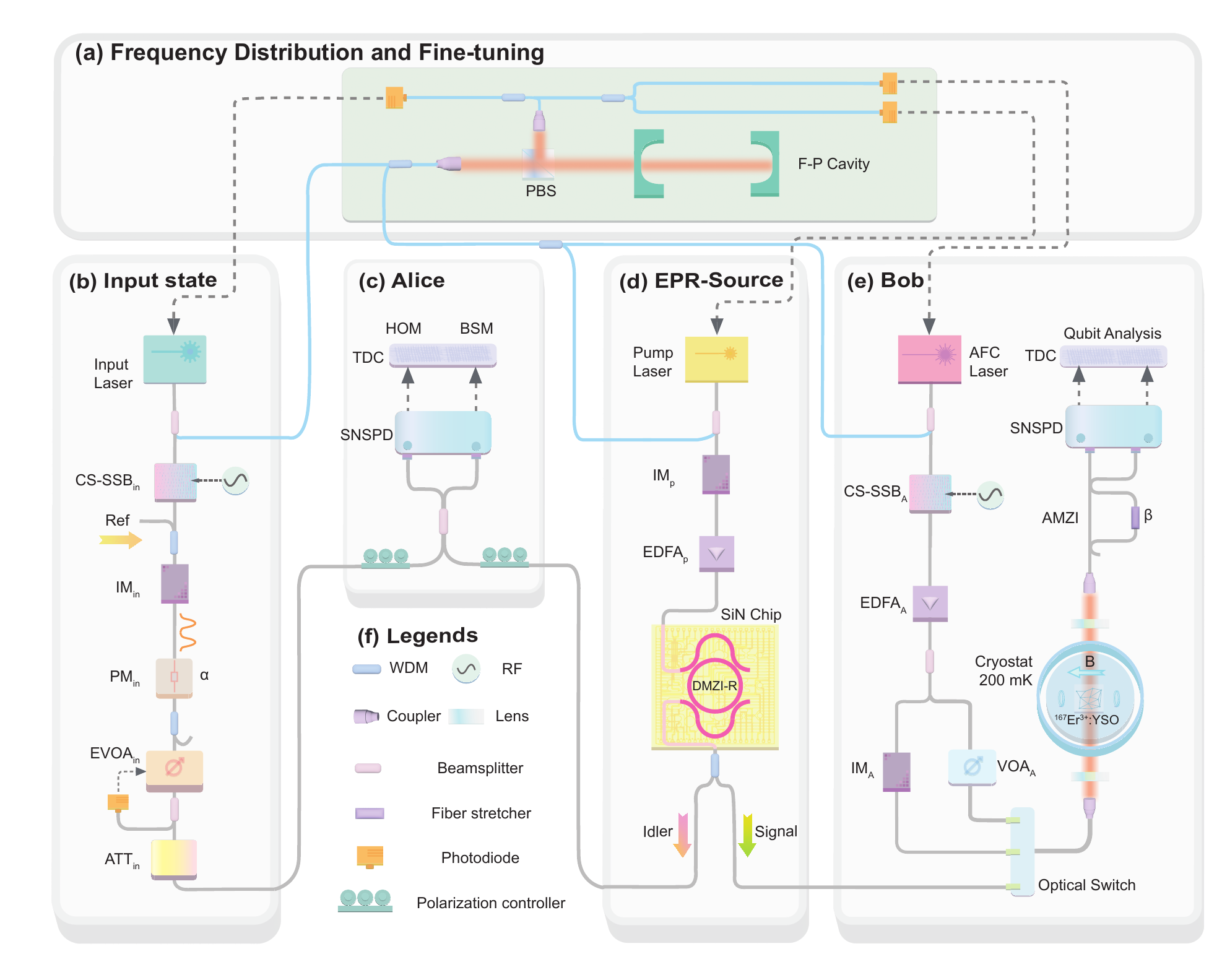}
      	\caption{\label{fig1} Schematics of the experiment setup. (a) Frequency distribution and fine-tuning module facilitates frequency and phase stability in this work. (b) The preparation of input state for quantum teleportation. The frequency, pulse shape and phase of the input photons are controlled with a carrier-suppression single-side-band modulator (CS-SSB${\rm {}_{in}}$), an intensity modulator (${\rm IM_{in}}$) and a phase modulator (${\rm PM_{in}}$), respectively. (c) Hong-Ou-Mandel interference or Bell-state measurement performed by Alice. (d) EPR source for generating entangled photon pairs. (e) Quantum memory by Bob. (f) Acronyms of the experimental components. Electrical variable optical attenuator (EVOA), attenuator (ATT), time-to-digital converter (TDC), superconducting nanowire single-photon detector (SNSPD), polarization beam splitter (PBS), wavelength division multiplexing (WDM), radio-frequency (RF), erbium-doped fiber amplifier (EDFA), dual Mach-Zehnder interferometer micro-ring (DMZI-R), asymmetric Mach-Zehnder interferometer (AMZI). See text for details.}
	\end{figure*}

    The schematic of our experimental setup is shown in Fig. \ref{fig1}. The quantum teleportation from telecom photons to erbium ions is realized by implementing the following steps: (1) Frequency distribution and fine-tuning (Fig. \ref{fig1}(a)): this work requires frequency and phase stability across about 600-GHz span, corresponding to the frequency difference between the signal photon and the idler photon, which is realized by frequency locking three different frequency light with different modes of a single Fabre-P\'{e}rot (F-P) cavity;  See Note S2 of Ref.\cite{SI} for details on frequency distribution and fine-tuning. (2) Input state preparation (Fig. \ref{fig1}(b)): we prepare the input photons which carried the states need to be teleported from attenuated laser locked to F-P cavity. The input states can be written as:
    \begin{equation}
		\label{eq1}
  		|\psi\rangle_{\rm in}=\alpha|{\rm e}\rangle+\beta e^{i\varphi}|{\rm l}\rangle,
    \end{equation}
    where $|{\rm e}\rangle$ and $|{\rm l}\rangle$ represent early and late temporal modes and $\varphi$ represents the relative phase between the two time-bins, and $\alpha^2+\beta^2=1$.
    (3) Entangled photon pair generation from EPR source (Fig. \ref{fig1}(d)): we use a pulsed pump laser to excite the dual Mach-Zehnder interferometer micro-ring (DMZI-R) resonator to create a pair of entangled (signal and idler) photons at the telecom C-band and narrow bandwidth compatible with ${\rm ^{167}Er^{3+}}$ optical transition.  We generate the time-bin entangled state with Franson scheme\cite{franson_PRL_1989} as
    \begin{equation}
		\label{eq2}
  		|\Phi^{+}\rangle_{\rm i,s}=\frac{|{\rm ee}\rangle_{\rm i,s}+|{\rm ll}\rangle_{\rm i,s}}{\sqrt{2}},
	\end{equation}
    where ``i'' and ``s'' denote the ``idler'' and ``signal'' photons.
    (4) Bell-state measurement by Alice (Fig. \ref{fig1}(c)): the three-photon state here can be written as a joint state $|\Psi\rangle=|\psi\rangle_{\rm in}\otimes|\Phi^{+}\rangle_{\rm i,s}$. We interfere the idler photon and the input photon on a fiber beam splitter (BS) to perform BSM using Hong-Ou-Mandel (HOM) two-photon interference\cite{Weinfurter_1994}.   {By doing so, we project them on to Bell-state $|\Psi^{-}\rangle=\tfrac{1}{\sqrt{2}}(|{\rm el}\rangle-|{\rm le}\rangle)$, heralding the quantum state of the signal photons to $|\psi_{\rm s}\rangle=-\sigma_{y}|\psi_{\rm in}\rangle$, where $\sigma_{y}$ is the Pauli-Y matrix.} (5) Quantum memory (Fig. \ref{fig1}(e)): we send the signal photon into the erbium ions to store for more than 2 $\mu$s; and then analyze the quantum state fidelity of the retrieved signal photon to the input photon with an asymmetric Mach-Zehnder interferometer (AMZI) and verify the successful quantum teleportation from telecom photons to erbium ions.
    
\section{Input state preparation}

	As shown in Fig. \ref{fig1}(b), to generate the time-bin qubit to be teleported to Bob, the output of a 1541 nm continuous-wave (CW) laser (Input Laser) is chopped by an intensity modulator (${\rm IM_{in}}$) into one pulse (for the pole states) or two pulses separated by 32 ns (for the equator states) with a pulse duration of 4 ns and a repetition rate of 10 MHz. Subsequently, a phase modulator (${\rm PM_{in}}$) is employed to control the relative phase of the two adjacent pluses to encode the quantum state of the input photons. We further use an electrical variable optical attenuator (EVOA${\rm{}_{in}}$) to reduce the mean photon number to the single-photon level, $\mu\simeq0.0825$. A $9:1$ fiber BS combined with a photodiode is used to provide feedback to the EVOA${\rm{}_{in}}$ to stabilize the mean photon number of the input photons. A carrier-suppression single-side-band modulator (CS-SSB${\rm{}_{in}}$) controls the central frequency of input photons. See Note S3 of Ref.\cite{SI} for details on phase stabilization of the input state, and Note S4 for optimizing the temporal delay of ${\rm PM_{in}}$.
    
\section{Entangled photon-pair source}
    By leveraging advances in complementary metal-oxide-semiconductor fabrication techniques, integrated quantum photonics is a promising approach for realizing the source of entangled photons. Recent progress has focused on the generation of energy-time entangled optical frequency combs within a resonator\cite{Reimer_OE_2014,Mazeas_OE_2016,Jaramillo-Villegas_Optica_2017,Samara_OE_2019,Zenghong_PRL_2024}. Based on our previous work, we employ a SiN MRR to generate comb-like bipartite energy-time entangled states through the nonlinear process of cavity enhanced SFWM \cite{Samara_OE_2019,Li-Xiaoying_PRL_2005,ramelow2015silicon_nitride_platform,Samara_2021,Fan_lpor_2023,Liu_junqiu_PRL_2024}. The time-bin entangled photon pairs are generated from an integrated SiN\cite{Jaramillo-Villegas_Optica_2017,Samara_OE_2019,Li-Xiaoying_PRL_2005,ramelow2015silicon_nitride_platform,Samara_2021,Wenjun_PRApplied_2022,kues_quantum_2019,lu_chip-integrated_2019,Wenjun_PRApplied_2023} DMZI-R resonator\cite{Vernon_OL_2017,Tison_OE_2017,lu_three-dimensional_2020,wu_bright_2019,Leizhen_PRL_2023}. A frequency-locked 1538 nm CW laser in conjunction with ${\rm IM_{p}}$ is employed to generate the pump pulses. Two pulses of 4 ns duration are separated by 32 ns at a repetition rate of 10 MHz, matching the time sequence of the input photons. These pump pulses are then amplified to an average on-chip pump power of 1.35 mW with an erbium-doped fiber amplifier (EDFA$_{\rm p}$). A pair of signal ($\sim$1536 nm) and idler ($\sim$1541 nm) photons at telecom-wavelength with narrow bandwidth ($\sim$185 MHz) is generated from the resonator.

\section{Indistinguishability of photons at the BSM}
\begin{figure}[!htbp]
  	\includegraphics[width=1\linewidth]{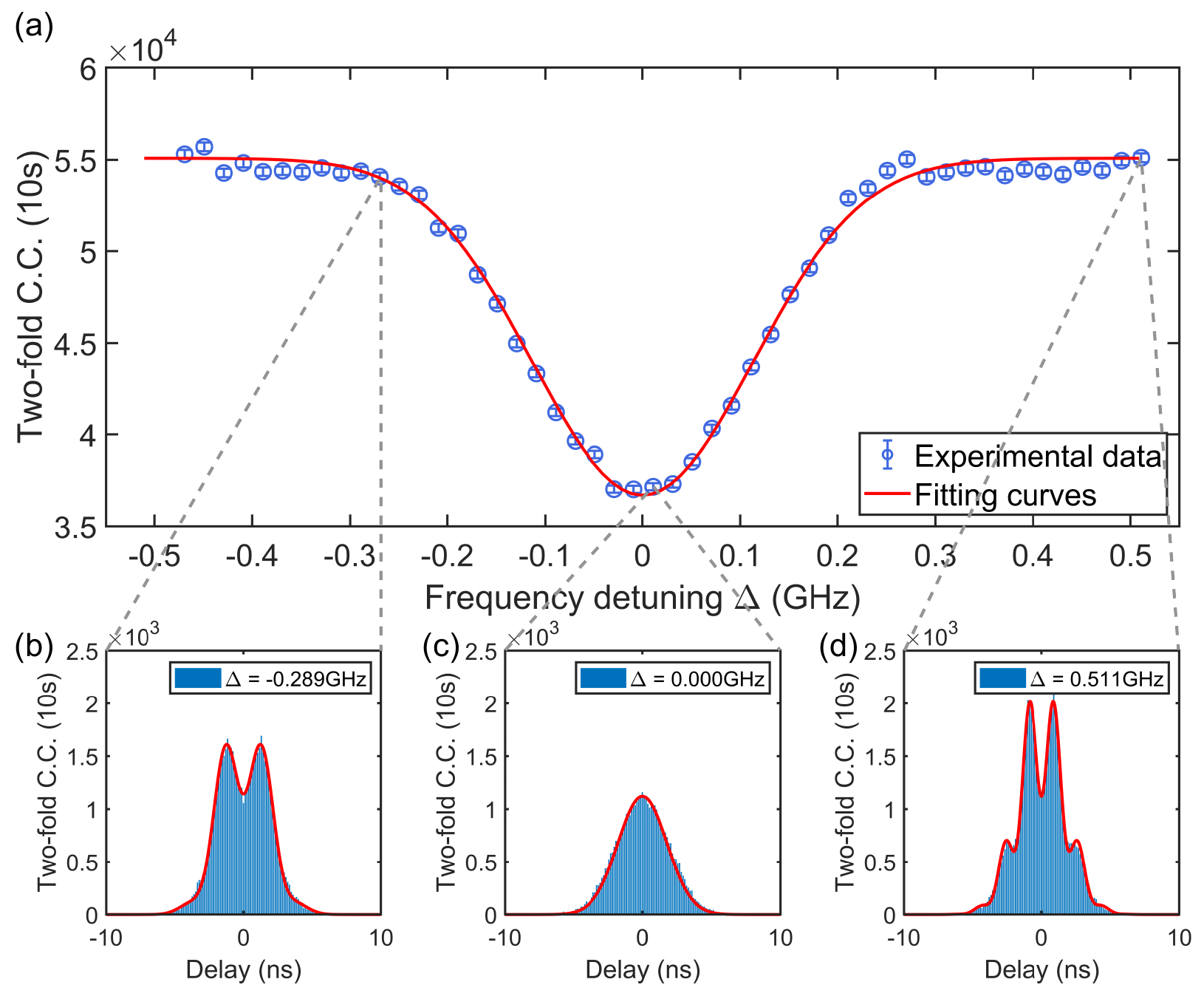}
  	\caption{\label{fig2} Hong-Ou-Mandel interference in the frequency domain and time resolved two-photon beating interference results. (a) Two-photon coincidence counts as a function of relative frequency detuning between the input and idler photons. We firstly overlap the temporal and spatial modes of the input and idler photons on the fiber beam splitter. Then we fix the frequency of idler photon and adjust the central frequency of the input photon with a CS-SSB${\rm _{in}}$. Destructive interference shows up when two photons’ frequency are aligned. The time resolved two-photon beating interference results for the frequency differences at $-0.289$ GHz (b), 0 GHz (c) and 0.511 GHz (d), respectively. The experimental data for frequency dependent beating fringes are shown in blue bars and the red curves are the theoretical fits.}
  	\end{figure}

    For a successful BSM, we need to ensure indistinguishability between input photons and idler photons in all degrees of freedom, including spatial, temporal, polarization and spectral modes. The indistinguishability of polarization- and spatial- mode is ensured by polarization controllers and single-mode fibers. By precisely controlling the electrical delay of the ${\rm IM_{in}}$'s drive signal, the input photons and idler photons arrive simultaneously at the BS. See Note S5 of Ref.\cite{SI} for details. We estimate the indistinguishability of two photons using HOM quantum interference\cite{HOM_PRL_1987} in the frequency domain. CS-SSB${\rm _{in}}$ adjusts distinguishability using the relative detuning $\Delta$ between the central frequency of the input photon and the idler photon. Two-fold coincidence counts (6 ns coincidence window) between two superconducting nanowire single-photon detectors (SNSPDs) placed at the output ports of the BS are measured for various $\Delta$, with results shown in Fig. \ref{fig2}. The visibility of the HOM-dip is $33.5\pm0.5\% $, corresponding to an indistinguishability of $83.7\pm1.1\%$. Here, the upper bound of the visibility is 40\% for the ideal interference of the coherent state and the thermal state\cite{photonics_2023}.   {The main factors that limit the HOM visibility are spectral lineshape and linewidth of input and idler photons. The deviation of HOM visibility from the ideal will lead to an increase in the quantum bit error rate (QBER) of X- and Y-basis but not affect the QBER of Z-basis\cite{valivarthi_quantum_2016}. See Note S6 of Ref.\cite{SI} for details.}

    When the central frequencies of the input photon and the idler photon are not the same and the frequency detuning $\Delta \ne 0$, a time-resolved quantum beat signal\cite{QBeating_Ou_Mandel_1988,twocolor_Rarity_1990,legero_time-resolved_2003,photonics_2023} with a frequency detuning of $\Delta$ in the interference appears, as shown in Fig. \ref{fig2}(b) $\Delta=-0.289$ GHz, (c) $\Delta=0$ GHz, (c) $\Delta=0.511$ GHz, respectively. See Note S6 of Ref.\cite{SI} for details on theoretical models for HOM interference and quantum beating.

\section{Quantum memory based on erbium ion ensemble}
    Our quantum memory is prepared in a 50 ppm doped ${\rm ^{167}Er^{3+}\!\!:\!\!Y_{2}SiO_{5}}$ crystal using the atomic-frequency comb (AFC) protocol\cite{AfzeliusPRA2009},  which offers an inherent temporally multiplexed storage ability. The crystal with dimensions of ${\rm 4\times5\times9\ mm^3}$ (${D_1, D_2, b}$) sits in a dilution refrigerator at about 200 mK with the light propagating along the ${b}$ axis. Moreover, the crystal is mounted with a rotational stage to align the ${D_1}$ axis of the crystal at an angle of about 120${{}^{\circ
    }}$ with the 1.5-T magnetic field. This experimental condition enlarges the electronic Zeeman level’s splitting of ${\rm {}^{167}Er^{3+}}$ so as to freeze the electron spin to improve the ions’ lifetime and coherence time \cite{Rancic_coherence_2018, bottgerPRB_2006}.  To store the signal photons, we prepare a 200-MHz AFC with a period of 0.457 MHz, corresponding to a storage time of 1/0.457 MHz = 2187 ns, by tailoring the absorption spectrum of the $^4I_{15/2}(0)\ ( m_{S}=-1/2)\leftrightarrow {}^4I_{13/2}(0)\ (m_{S}=-1/2)$ transition of 1536 nm. The overall storage efficiency is about 1.1\%. See Note S7 of Ref.\cite{SI} for details on the quantum memory and experimental time sequence.   

\section{Quantum teleportation results}
	Following the generation of entangled photon pair, the signal photons are sent to the quantum memory. Meanwhile, the input photons and the idler photons are sent to a $50:50$ fiber BS to perform the Bell-state measurement by detecting them by two SNSPDs at each output of the BS, with their arrival times recorded using a time-to-digital converter. One of the two detectors triggered in the early time bin and the other in the late time bin corresponds to a projection into the Bell-state $|\Psi^{-}\rangle$. See Note S1 of Ref.\cite{SI} for details. After approximately 2187 ns storage time, Bob retrieves the signal photon and analyzes its quantum state using an AMZI for the equatorial input states. For the pole input states, Bob removes the AMZI and directly detects the arrival time of the retrieved photons for the projections into early and late states.

    To obtain complete information about the retrieved states, we perform quantum state tomography (QST) to reconstruct the density matrices $\rho$ for the quantum states after teleportation\cite{Daniel_PRA_2001,Photonic_ST_2005,Takesue_Optica_2015}. The retrieved states after teleportation are measured by quantum state fidelities, $F=\langle\psi_{\rm B}|\hat{\rho}_{\rm out}|\psi_{\rm B}\rangle$ with the ideal states $|\psi_{\rm B}\rangle$ for the four teleported states. The fidelity results are listed in Table. 1. The average fidelity of an arbitrary state is $\bar{F}=\frac{1}{6}(F_{\rm e}+F_{\rm l}+2F_{\rm +}+2F_{\rm +i})$ and is about $0.818\pm0.019$ in our work, which is more than seven standard deviations above the classical bound of 2/3\cite{MassarPRL1995}. We note that the 2/3 bound is only applied if the input state is encoded into genuine single photons. Here we use the decoy state method (DSM)\cite{Decoy_PRL_2005,ma_practical_2005,valivarthi_quantum_2016} to calculate the lower bound of single-photon teleportation fidelity $F^1_{\text{Lower}}$ based on all-optical setup, which is $81.82\pm1.25\%$. Our results surpass the classical bound by more than 12 standard deviations. See Note S9 of Ref.\cite{SI} for more details on quantum state tomography and Note S12 for more details on DSM.

    \begin{table}[!htbp]
	\begin{ruledtabular}
	\begin{tabular}{cccc}
    	Input state  &  Expected state  &   Fidelity   &   Average fidelity  \\
    	\hline
    	$|{\rm e}\rangle$   &  $|{\rm l}\rangle$  &  0.840$\pm$0.010  &   \multirow{4}{*}{0.818±0.019}   \\
        $|{\rm l}\rangle$	  &  $|{\rm e}\rangle$  &  0.879$\pm$0.009  &      \\
    	$|{\rm +}\rangle$   &  $|{\rm -}\rangle$  &  0.794$\pm$0.023  &      \\
        $|{\rm +i}\rangle$  &  $|{\rm +i}\rangle$ &  0.800$\pm$0.024  &      \\
  	\end{tabular}
   
	\end{ruledtabular}
    \caption{\label{table1}Experimental quantum state fidelities obtained from density matrices reconstructed with quantum state tomography measurement. The uncertainties in state fidelities extracted from these density matrices are calculated using a Monte Carlo routine assuming Poissonian errors.}
	\end{table}
    
\begin{figure}[!htbp]
  	\includegraphics[width=1\linewidth]{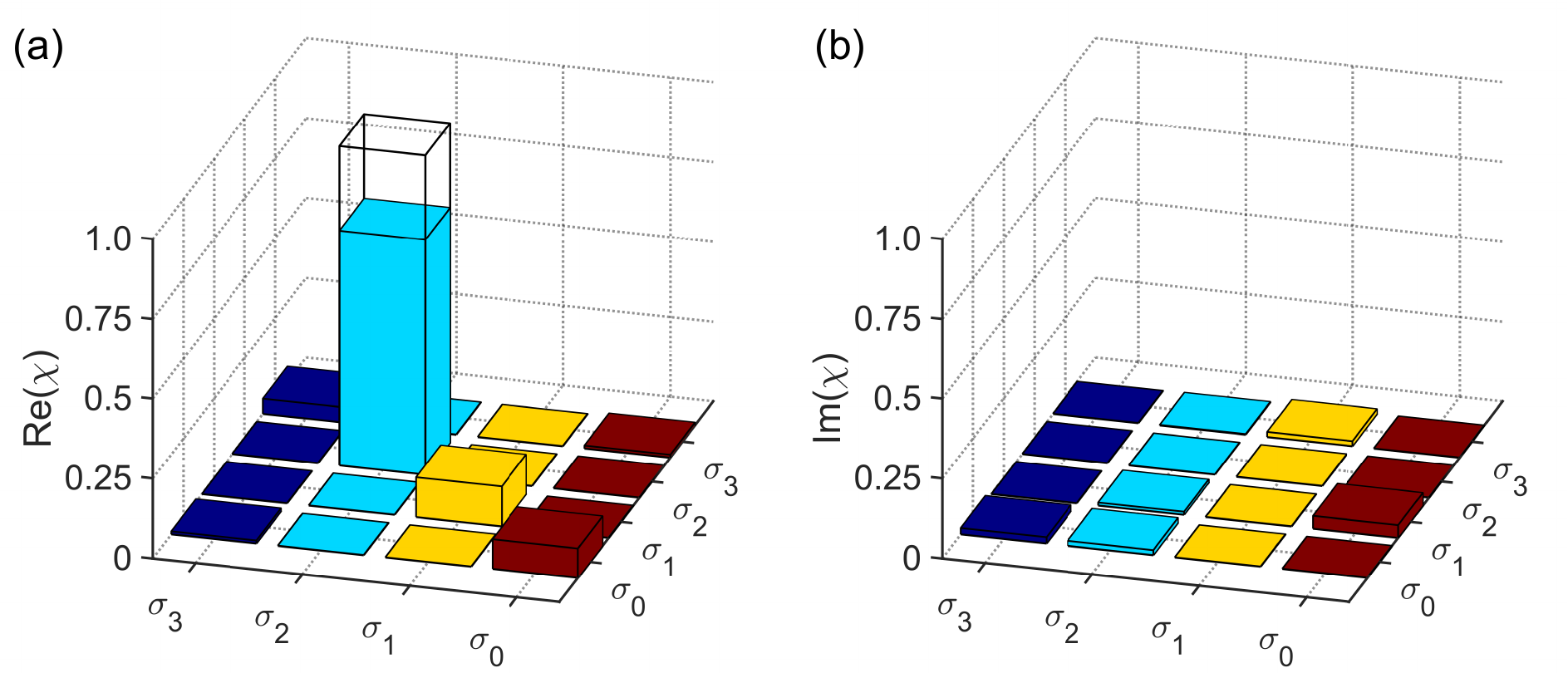}
  	\caption{\label{fig3} Quantum process tomography of quantum teleportation. (a) The real and (b) the imaginary parts of the reconstructed quantum process matrix. The bar graphs show the reconstructed real and imaginary parts of the process matrices for our experimental quantum teleportation process. The wire grids indicate the expected values for the ideal cases.}
	\end{figure}

    To fully quantify the quantum teleportation process, we perform quantum process tomography (QPT) to determine the process matrix $\chi$, defined by 
    \begin{equation}
        \label{eq4}
        \rho=\sum^{3}_{l,k=0}\chi_{lk}\sigma_{l}\rho_{\rm ideal}\sigma_k.
	\end{equation}
    Here, the $\sigma_i$ are the Pauli matrices with $\sigma_{0}$ the identity operator. $\rho_{\rm ideal}$ corresponds to four input states $|{\rm e}\rangle\langle{\rm e}|$, $|{\rm l}\rangle\langle{\rm l}|$, $|{\rm +}\rangle\langle{\rm +}| $ and $ |{\rm +i}\rangle\langle{\rm +i}|$, respectively. Fig. \ref{fig3}(a) and \ref{fig3}(b) show the real and imaginary parts of the process matrix $\chi$. The output states of the signal photons after teleportation are the input states up to a unitary operation $\sigma_{y}$. Therefore, the ideal process matrix of our experiment has only one non-zero component, $(\chi_{\rm ideal})_{33}$= 1. See Note S1 of Ref.\cite{SI} for details. The quantum process fidelity is $f_{\rm process}={\rm Tr}(\chi_{\rm ideal}\chi)=0.736\pm0.022$, which is more than ten standard deviations above the maximum process fidelity of 0.5 for the classical strategy.

\section{Conclusions}
    We have demonstrated the quantum state teleportation from telecom photons to a solid-state quantum memory based on erbium-ion ensembles. Using QST and QPT methods, we measure an average quantum state fidelity of 81.8\% and a quantum process fidelity of 73.6\%, respectively, which surpass the classical bounds by more than seven standard deviations.  We multiplex a F-P cavity to realize frequency distribution and fine-tuning and increase the stabilization of the whole experiment. 

    However, in order to realize practical applications, several aspects of our experiment need to be further improved. Due to the remaining temporal and spectral distinguishability of the input photons and idler photons, the HOM visibility we measured is lower than the ideal case. Although higher SFWM pump power is beneficial to enhance the counting rates, multi-photon events further reduce the teleportation fidelity. To circumvent the intrinsic trade-off between entangled photon rates and entanglement fidelity, one may consider using either on-demand entangled photon pair sources\cite{PhysRevLett.126.140501} or entanglement filter\cite{ye_photonic_2023} in the future. Moreover, using the SNSPDs with short recovery time allows us to detect two of four Bell-states and one can always get the same teleported states with active feed-forward. To extend the teleportation distance, quantum memory with longer storage time is necessary. One could use spin-wave AFC protocol to realize a millisecond storage and on-demand readout, which improves the entanglement distribution rate in quantum networks\cite{Spectral_Mulplex_PRL_2014,lago-rivera_telecom-heralded_2021,hänni2025,PRL_Simon_2007,RMP_2011}. Comparing with ${\rm ^{171}Yb^{3+}\!\!:\!\!Y_{2}SiO_{5}}$\cite{ortu_simultaneous_2018,Yb_spinwaveQM_PRL2020}, the level structure of ${\rm ^{167}Er^{3+}\!\!:\!\!Y_{2}SiO_{5}}$ is more complicated and requires more complicated optical pump and coherent control pulse sequences. In comparison with the efficiency of 18.8$\%$ at 10-$\mu$s storage in prior work performing quantum teleportation with ${\rm Pr^{3+}}$\cite{lago-rivera_long_2023}, our two-level AFC efficiency is lower mainly due to the near 1/e-reduction for the residual background absorption (See Note S7 of Ref.\cite{SI} for details) and a lower doping concentration. Other than the above aspects, we need to address the efficiency fluctuations that can occur due to the mount vibrations and magnetic field strength drift. Besides, cavity-enhanced AFC  approaches 100\% efficiency in principle needs to be employed, which has been demonstrated to be very promising\cite{Sabooni_2013,Cavity_AFC_PRL_2013,Jobez_2014,Cavity_AFC_PRA_2020,Duranti_2024}.

    With a quantum memory at telecom-wavelength, three photons in our experiment are all at telecom-wavelength, making it feasible to extend the teleportation distance based on standard optical fiber between Alice and Bob. It is also important for long-distance quantum communication based on quantum repeaters. Our work verifies the maturity and applicability of such technologies in quantum networks.

\section{Acknowledgment}
 This research was supported by the National Key Research and Development Program of China (Grants Nos. 2022YFE0137000, 2019YFA0308704), the Natural Science Foundation of Jiangsu Province (Grants Nos. BK20240006, BK20233001), the Leading-Edge Technology Program of Jiangsu Natural Science Foundation (Grant No. BK20192001), the Innovation Program for Quantum Science and Technology (Grants Nos. 2021ZD0300700 and 2021ZD0301500), the Fundamental Research Funds for the Central Universities (Grants Nos. 2024300324), and Nanjing University-China Mobile Communications Group Co.,Ltd. Joint Institute.

    
    \onecolumngrid
    \setcounter{section}{1} 
    \setcounter{figure}{0} 
    \setcounter{equation}{0}
    \setcounter{table}{0}
	\renewcommand*{\thefigure}{S\arabic{figure}}
    \renewcommand*{\thetable}{S\arabic{table}}
    \renewcommand*{\thesection}{S\arabic{section}}
    \renewcommand*{\theequation}{S\arabic{equation}}
	\counterwithout{equation}{section}
    \numberwithin{equation}{section}

\newpage
\section*{Supplemental Materials}
\section{Note S1: Quantum teleportation protocol}
	In this section we will briefly outline the teleportation protocol\cite{PhysRevLett.70.1895_1993}. Before performing the quantum teleportation experiment, we prepare the following quantum states: 

    \textbf{Entangled state}: The time-bin entangled photon pairs are generated from an integrated silicon nitride (SiN) dual Mach-Zehnder interferometer microring resonator (DMZI-R), using the spontaneous four-wave mixing (SFWM) process. The time-bin entangled state is
    \begin{equation}
		\label{eqS1}
  		|\Phi^{+}\rangle_{\rm i,s}=\frac{|{\rm ee}\rangle_{\rm i,s}+|{\rm ll}\rangle_{\rm i,s}}{\sqrt{2}},
	\end{equation}
    where s and i denote the ‘signal’ and ‘idler’ photons, and $|{\rm e}\rangle$ and $|\rm l\rangle$ represent the early and late temporal modes of the single photons, respectively. We set the temporal separation between $|{\rm e}\rangle$ and $|\rm l\rangle$ to be 32 ns. Each bin has a duration of approximately 4 ns.

    \textbf{Input state}: Using an intensity modulator and a phase modulator, we can generate an arbitrary qubit state in the time basis by modulating the amplitude and phase of weak coherent state. The input state can be written as
    \begin{equation}
		\label{eqS2}
  		|\psi\rangle_{\rm A}=\alpha|{\rm e}\rangle+\beta e^{i\varphi}|{\rm l}\rangle,
	\end{equation}
    where $\varphi$ is the relative phase between the two temporal modes, $\alpha$ and $\beta$ are real, and $\alpha^2+\beta^2=1$.
    
    Three-photon state mentioned above is
    \begin{equation}
		\label{eqS3}
  		|\Psi\rangle_{\rm Ais}=|\psi\rangle_{\rm A}\otimes|\Phi^{+}\rangle_{\rm i,s}.
	\end{equation}
    We can now rewrite this joint state in terms of the four maximally entangled Bell-states of the input and the idler photons: 
    \begin{equation}
		\label{eqS4}
  		|\Psi\rangle_{\rm A}=|\psi\rangle_{\rm A}\otimes|\Phi^{+}\rangle_{\rm i,s}=\frac{1}{2}
    \begin{bmatrix}
        |\Phi^{+}_\text{A,i}\rangle(\alpha|{\rm e}\rangle+\beta e^{i\varphi}|\textrm{l}\rangle)_{\textrm{s}}    \\
        +|\Phi^{-}_\textrm{A,i}\rangle(\alpha|{\rm e}\rangle-\beta e^{i\varphi}|\textrm{l}\rangle)_{\textrm{s}}    \\
        +|\Psi^{+}_\textrm{A,i}\rangle(\beta e^{i\varphi}|{\rm e}\rangle+\alpha |\textrm{l}\rangle)_{\textrm{s}}    \\
        -|\Psi^{-}_\textrm{A,i}\rangle(\beta e^{i\varphi}|{\rm e}\rangle-\alpha |\textrm{l}\rangle)_{\textrm{s}}
    \end{bmatrix}
	\end{equation}
    with the Bell-states $|\Phi^{\pm}_\text{A,i}\rangle=\tfrac{1}{\sqrt{2}}(|\textrm{ee}\rangle_\textrm{A,i}\pm|\textrm{ll}\rangle_\textrm{A,i})$ and $|\Psi^{\pm}_\textrm{A,i}\rangle=\tfrac{1}{\sqrt{2}}(|\textrm{el}\rangle_\textrm{A,i}\pm|\textrm{le}\rangle_\textrm{A,i})$.

    By performing Bell-state measurement (BSM), the probability of projecting input photon and the idler photon onto one of the Bell-states is 1/4. In our scheme, we identify the Bell-state $|\Psi^{-}\rangle_\textrm{A,i}$. Hence, the signal photon is projected onto the input quantum state up to a $\sigma_y$ operation, i.e. a bit flip and a subsequent phase flip operation.
    
\begin{figure}[h]
	\includegraphics[width=0.8\linewidth]{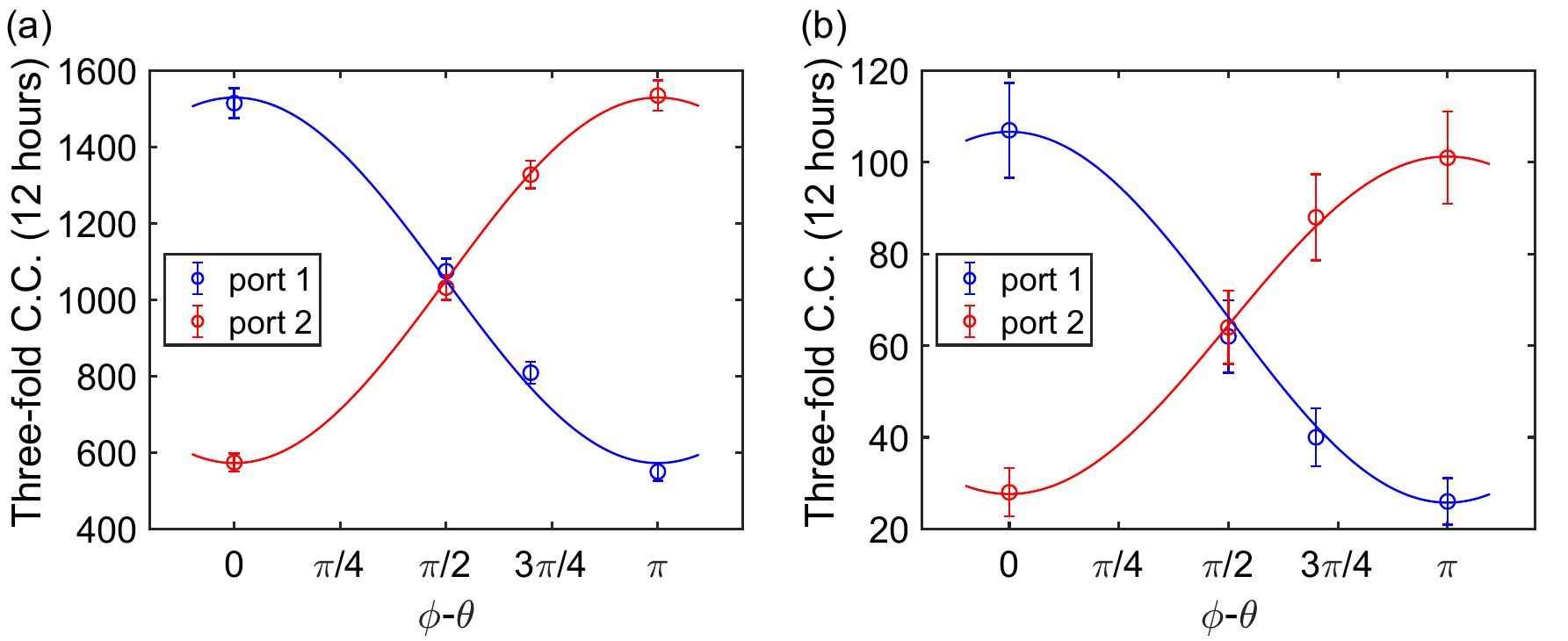}
      	\caption{\label{figS12} Three-fold coincidence counts for (a) transmission and (b) echo photons in the two output ports of Bob's analyzing AMZI. To show the coherence of quantum teleportation process, we vary the phase of the input state, $\varphi$, in the basis that is rotated around the equator of the Bloch sphere and fix the analyzer phase, $\theta$. Solid lines show the respective theoretical fits.}
	\end{figure}

    \textbf{Expected state measurement}: After a successful BSM, the state of the signal photon becomes $|\psi\rangle_\textrm{s}=\beta e^{i\varphi}|\textrm{e}\rangle-\alpha|\textrm{l}\rangle$, with $\alpha=\beta=\tfrac{1}{\sqrt{2}}$. To characterize the fidelity of the state of signal photons after teleportation, Bob prepares an asymmetric Mach-Zehnder interferometer (AMZI) as the time-bin qubit analyzer. The two output ports of Bob’s AMZI correspond to projections onto the states $|\psi^{\pm}\rangle=\tfrac{1}{\sqrt{2}}(|\textrm{e}\rangle\pm e^{i\theta}|\textrm{l}\rangle)$. Hence, the probability of three-fold coincidence counts in the two output ports can be written as
    \begin{equation}
		\label{eqS5}
  		|\langle\psi^{\pm}|\psi\rangle_\textrm{s}|^2=\frac{1}{2}(1\pm\cos(\varphi-\theta)),
	\end{equation}
    where $\varphi$ is the phase of Alice's input photon and $\theta$ is the phase of Bob's analyzing AMZI. The experimental results are shown in Fig.  \ref{figS12}.

\section{Note S2: Frequency stabilization, dissemination and translation}
    For a successful BSM, the input photon and idler photon need to be indistinguishable at the beam splitter. These two interfering photons have to be indistinguishable in all degrees of freedoms (DOFs), including polarization, spectral-, temporal- and spatial-modes. In our setup, there are three independent lasers with three different frequencies, the input laser (1540.9 nm), the pump laser (1538.5 nm) and the AFC laser (1536.17 nm). It is necessary to stabilize the frequencies of these three lasers. However, for commercial InGaAs APD, the frequency response is low comparing to the frequency differences of these three lasers. Therefore, it is challenging to stabilize the relative frequency between these three lasers by standard beat measurement. 
    
    \begin{figure}[H]
\centering
	\includegraphics[width=0.8\linewidth]{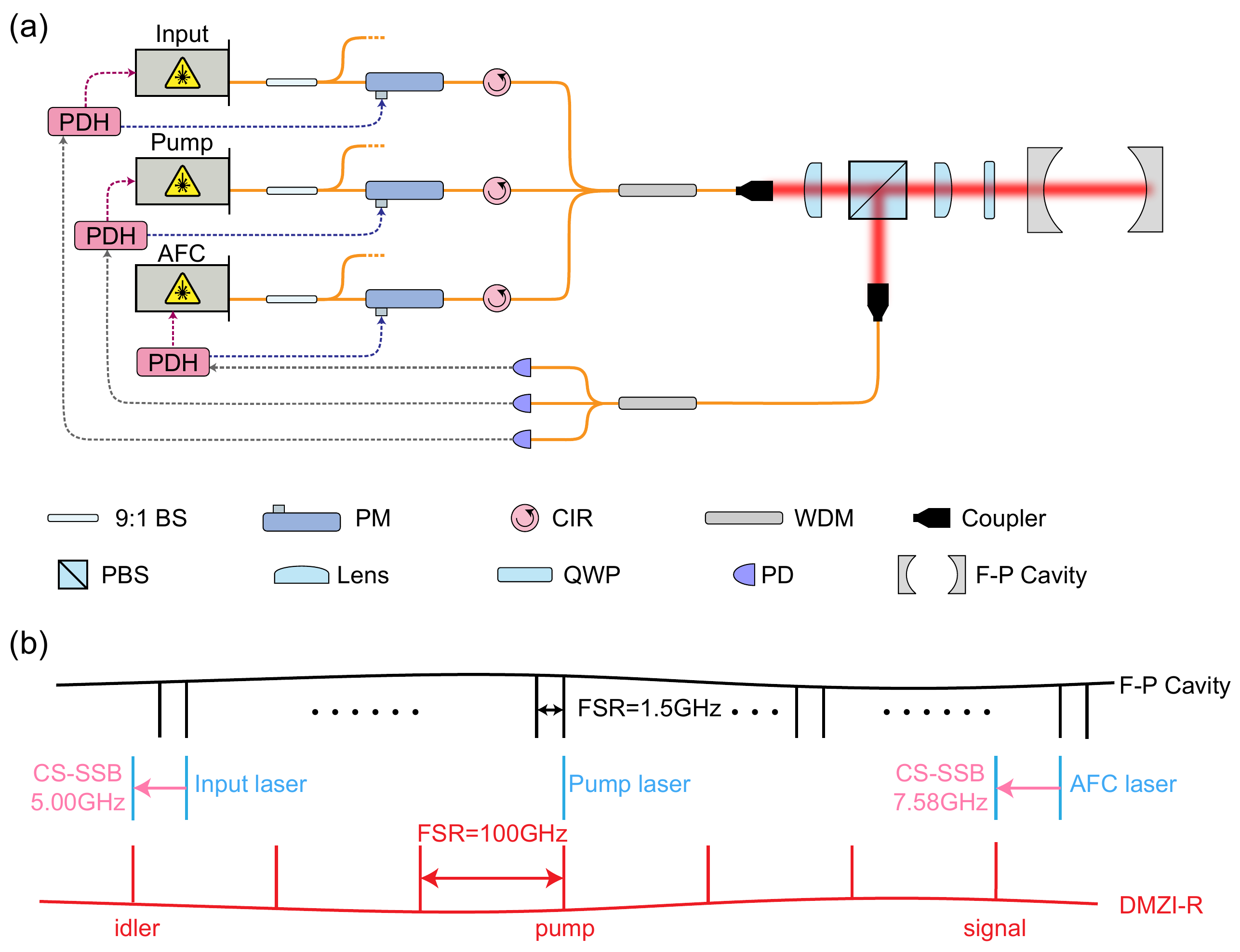}
      	\caption{\label{figS1}(a) Schematics of the experimental setup for multi-frequency locking and dissemination and translation. Legends: beam splitter (BS), phase modulator (PM), circulator (CIR), wavelength division multiplexing (WDM), polarization beam splitter (PBS), quarter-wave plate (QWP), Fabry-Pérot cavity (F-P cavity). (b) Frequency translation of various locked lasers. Black lines: the reflection spectrum of the F-P cavity; Red lines: the resonance frequency modes of the DMZI-R; Blue lines: frequency-locked lasers; Pink lines: CS-SSB${}_\textrm{in}$ and CS-SSB${}_\textrm{A}$.}
	\end{figure}
     
    The schematics of the experiment for multi-frequency stabilization is shown in Fig. \ref{figS1}(a). A fiber beam splitter (BS) with a ratio of $99:1$ is used to tap a fraction of the laser power for frequency locking. A phase modulator (PM) is employed to generate two sidebands with frequencies $\omega\pm\Omega$, where $\Omega=10$ MHz is the phase modulation frequency and $\omega$ is the carrier frequency. A three-port circulator (CIR) with more than 40 dB isolation is placed between the laser and the F-P cavity, preventing the back reflections that may cause intensity fluctuation and damage. Subsequently, several dense wavelength division multiplexing (DWDM) filters with 100 GHz spacing are cascaded to combine these three different frequencies and send to the F-P cavity. The linewidth of the F-P cavity is about 42 kHz and the free spectral range (FSR) of it is about 1.5 GHz. The reflections with error signals from the cavity are separated by the same cascaded multiple DWDMs and each detected respectively with a photodiode (PD). Three Pound-Drever-Hall (PDH) controllers are used to generate the electrical drive signals applied to the piezoelectric ceramic (PZT) of these lasers to stabilize the output frequencies.

    The conceptual scheme of frequency dissemination and translation are shown in Fig. \ref{figS1}(b). By locking the input, pump and AFC lasers to the F-P cavity modes (indicated with black lines in Fig. \ref{figS1}(b)), the frequency drift of the three lasers is reduced to less than 10 kHz (blue lines in Fig. \ref{figS1}(b)). The pump laser is then injected into the DMZI-R and generates a pair of photons, signal and idler (red lines in Fig. \ref{figS1}(b)). However, the frequency of the idler photon is not naturally aligned with the input laser, the same problem exists for the signal photon and the AFC laser. Here, we employ two carrier-suppression single-side-band modulator (CS-SSB${}_\textrm{in}$ and CS-SSB${}_\textrm{A}$) to translate and align the frequency of the input and AFC lasers independently (pink lines in Fig. \ref{figS1}(b)).

\section{Note S3: Phase stabilization}
    \begin{figure}[H]
    \centering
	\includegraphics[width=0.8\linewidth]{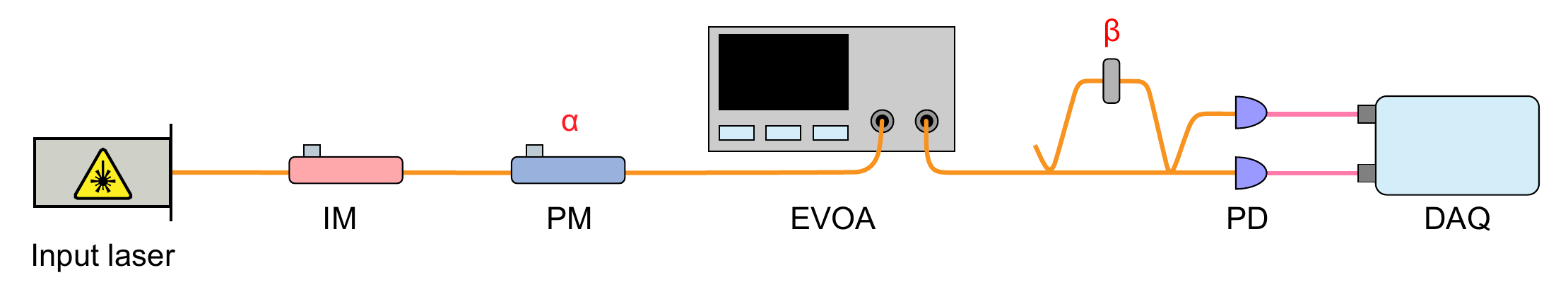}
      	\caption{\label{figS2} Schematics of the experimental setup for creation and measurement of time-bin qubit. }
	\end{figure}
    In the experiment, we prepare the time-bin qubit to be teleported and analyze its phase stability with the setup shown in Fig. \ref{figS2}. The input laser is intensity modulated into two pulses separated by 32 ns with a pulse duration of 4 ns and a period of 100 ns. A PM applies the relative phase $\alpha$, which can be expressed as
    \begin{equation}
		\label{eqS6}
  		\alpha=\frac{2\pi nL}{\lambda}=\frac{2\pi nLf_\textrm{input}}{c}
	\end{equation}
    between the early and late temporal modes, $|\textrm{e}\rangle$ and $|\textrm{l}\rangle$. $f_\textrm{input}$ is the frequency of the input laser and $L$ is the length of PM. Note that if we keep the input laser free-running, the frequency drift of the laser leads to the fluctuations in the relative phase of input state with respect to its analyzer.

    The phase of the analyzing AMZI is locked by PID feedback controllers using the transmission of the pump laser as a reference. The pump laser is locked to the F-P cavity using the setup shown in Fig. \ref{figS1}. Therefore, we can ensure that both the path-length difference $L_\textrm{AMZI}$ and the relative phase $\varphi_\textrm{ref}=2\pi nL_\textrm{AMZI}f_\textrm{ref}/c$ are constant.

    However, for an input photon whose frequency is independent of the pump laser, the relative phase of the AMZI becomes $\beta=\varphi_\textrm{input}=2\pi nL_\textrm{AMZI}f_\textrm{input}/c$, which is unstable due to the unlocked frequency $f_\textrm{input}$ of input laser.
    
\begin{figure}
	\includegraphics[width=1\linewidth]{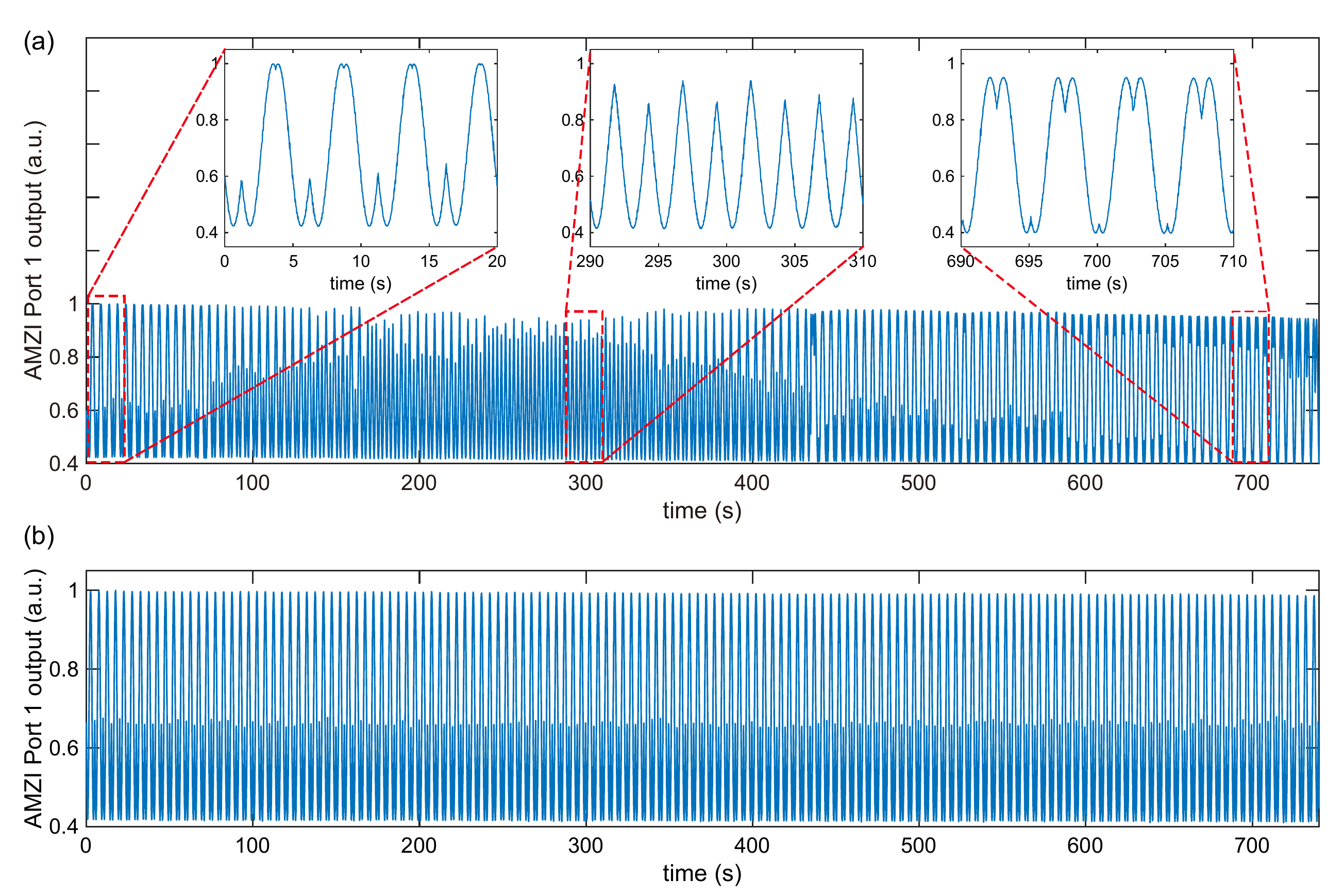}
      	\caption{\label{figS3}Comparison between the frequency locking being OFF versus ON for the input laser. (a) The input laser is free-running. (b) The input laser is locked.}
	\end{figure}

    To investigate the phase fluctuation caused by frequency drift of the input laser, we scan the PM voltage with a triangular periodic signal at a repetition rate of 0.2 Hz and detect the transmitted power with a PD at the output port of the AMZI. Note that after passing the AMZI, the PD can register optical pulses in three different time bins. The first (last) time bin corresponding to the early (late) pulse passing the short (long) arm of the AMZI and the middle time bin corresponding to the early (late) pulses passing the long (short) arm. Due to the low response bandwidth of PD, all of these events considered above are indistinguishable to PD's detection, leading to a 50\% visibility in principle. As shown in Fig. \ref{figS3}(a), the relative phase between $\alpha$ and $\beta$ fluctuates more than $\pi$ phase within a few hundred seconds when we keep input laser free-running. Conversely, this fluctuation disappeared if we lock the frequency of the input laser, as shown in Fig. \ref{figS3}(b).

    In the teleportation experiment, the signal photons will be analyzed by the AMZI instead of input photons. Therefore, the phase $\beta$ now can be rewritten as
    \begin{equation}
		\label{eqS7}
  		\beta=\varphi_\textrm{signal}=\frac{2\pi nL_\textrm{AMZI}f_\textrm{signal}}{c}.
	\end{equation}
    The frequency of the signal photon $f_\textrm{signal}$ is related with the frequency of the pump laser $f_\textrm{pump}$ by the FSR of the DMZI-R source:
    \begin{equation}
		\label{eqS8}
  		f_\textrm{signal}=f_\textrm{pump}+N_\textrm{s,p}\textrm{FSR}_\textrm{SiN},
	\end{equation}
    where $N_\textrm{s,p}=3$ corresponds to the relative frequency modes number between the signal mode,  and the pump mode of the DMZI-R. The FSR$_\textrm{SiN}$ is about 100 GHz. Hence, the phase $\beta$ can be rewritten as
    \begin{equation}\label{eqS9}
      \begin{split}
  		\beta=\varphi_\textrm{signal}=&\frac{2\pi nL_\textrm{AMZI}f_\textrm{signal}}{c} \\
        &=\frac{2\pi nL_\textrm{AMZI}}{c}\Big(f_\textrm{pump}+N_\textrm{s,p}\textrm{FSR}_\textrm{SiN}\Big) \\
        &=\varphi_\textrm{pump}+\frac{\varphi_\textrm{pump}}{f_\textrm{pump}}\times N_\textrm{s,p}\textrm{FSR}_\textrm{SiN}	
      \end{split}
    \end{equation}
    From the Eq. (\ref{eqS9}), we can get the relation between the fluctuation of the relative phase $\textrm{d}\beta$ and the frequency drift $\textrm{d}(f_\textrm{pump})$ of the pump laser:
    \begin{equation}
		\label{eqS10}
  		\textrm{d}\beta=-\varphi_\textrm{pump}N_\textrm{s,p}\textrm{FSR}_{\textrm{SiN}}\frac{\textrm{d}(f_\textrm{pump})}{(f_\textrm{pump})^2}.
	\end{equation}
    Here we assume that the FSR of the DMZI-R is constant. In our case, $\varphi_\textrm{pump}\approx$$ \tfrac{32\textrm{ns}\times2\times10^8\textrm{m/s}}{1536.17\textrm{nm}}$$=4.17\times10^6$, $N_\textrm{s,p}=3$, FSR${}_\textrm{SiN}\approx100$ GHz, $f_\textrm{pump}\approx195$ THz. Eventually, the relation is given by
    \begin{equation}
		\label{eqS11}
  		\textrm{d}\beta\sim3.28\times10^{-11}\times\textrm{d}(f_\textrm{pump}).
	\end{equation}
    The relative phase in the input state is given by
    \begin{equation}
		\label{eqS12}
  		\alpha=\frac{2\pi nL}{\lambda}=\frac{2\pi nLf_\textrm{input}}{c}.
	\end{equation}
    
    Then the relation between the fluctuation of the relative phase $\textrm{d}\alpha$ and the frequency drift d$(f_\textrm{input})$ of the input laser is given by, according to Eq. (\ref{eqS12}),
    \begin{equation}
		\label{eqS13}
  		\textrm{d}\alpha=\frac{2\pi nL}{c}\textrm{d}(f_\textrm{input})=4.62\times10^{-9}\times\textrm{d}(f_\textrm{input}),
	\end{equation}
    where $n$ corresponds to the refractive index of the PM and $L$ is the length of the PM. Comparing Eq. (\ref{eqS11}) with Eq. (\ref{eqS13}), we can see that the relative phase $\alpha$ requires more than two orders of magnitude higher frequency stability than $\beta$.
    \begin{figure}[h]
	\includegraphics[width=0.8\linewidth]{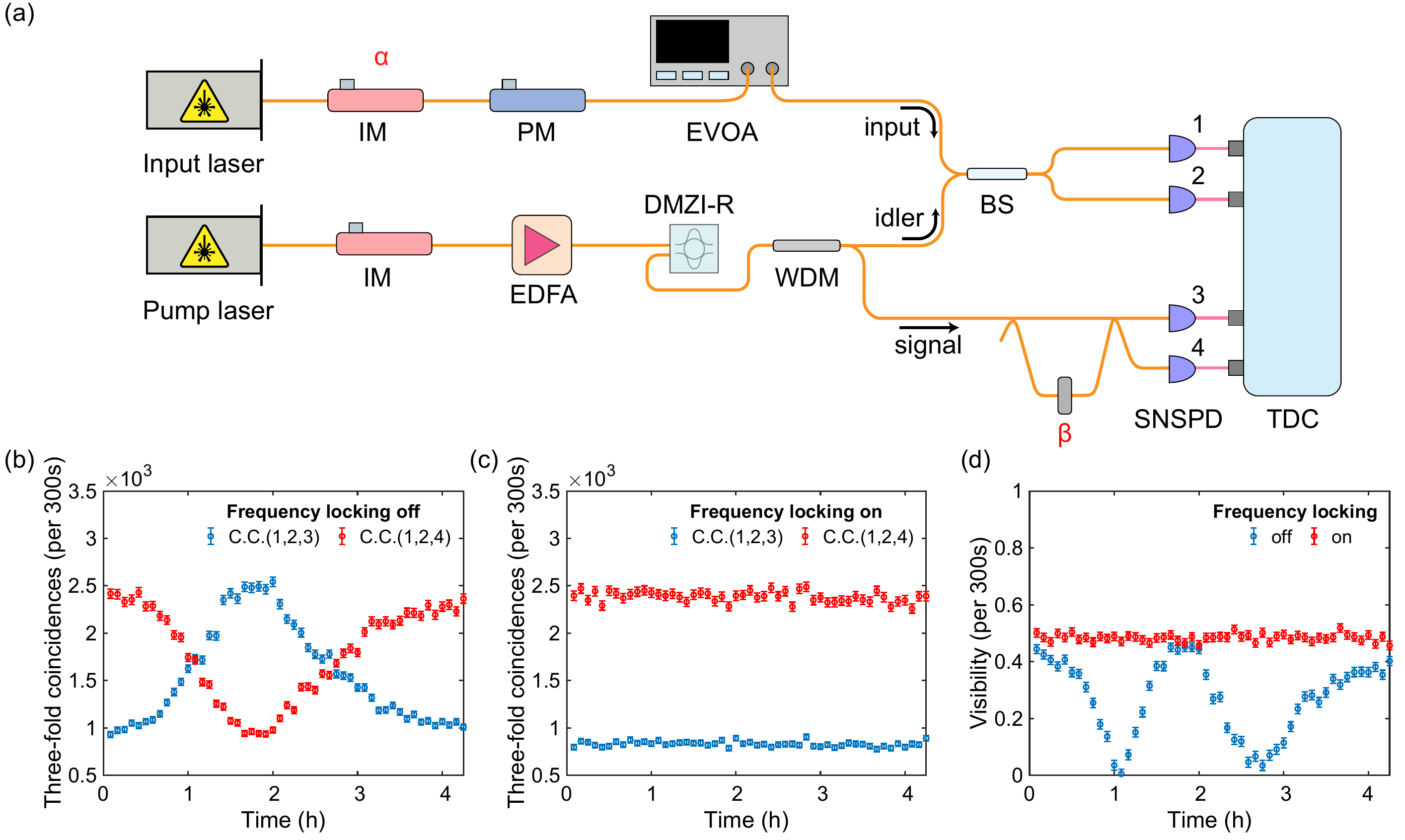}
      	\caption{\label{figS4}The phase stability of the three-fold coincidences and visibility. (a) All-optical experiment setup. (b) and (c) The three-fold coincidence results C.C.(1,2,3) and C.C.(1,2,4) as functions of integration time with frequency lock off and on. (d) The stability of visibility with frequency locking off and on.}
	\end{figure}

    In order to verify our model presented above, we use an all-optical experiment (without quantum memory) setup to investigate the frequency locking's impact on quantum teleportation, shown in Fig. \ref{figS4}(a). Firstly, we lock the frequency of pump laser and use it lock the phase of AMZI. Then, we either keep the frequency lock of the input laser off or on. We measure the three-fold coincidence C.C.(1,2,3) and C.C.(1,2,4) between input, idler and signal photons, where 1, 2, 3 and 4 represent four superconducting nanowire single-photon detectors (SNSPDs) placed at the output ports of the BS and the AMZI. The results as functions of integration time with frequency lock off and on are shown in Fig. \ref{figS4}(b) and (c), respectively. It is clear to see that, when locking is off, the three-fold coincidence counts undesirably drift away. On the contrary, when locking is on, the coincidence counts are kept stable. We show the visibility remain stable over several hours in Fig. \ref{figS4}(d). Based on these experimental results, we conclude our model for analyzing the phase and frequency relation between three lasers are correct. The measure we take can keep the relative phase ($\alpha-\beta$) stable enough within the integration time of our experiment.

\section{Note S4: Optimize the delay of PM${}_\textrm{in}$}
    We use the same setup as shown in Fig. \ref{figS2} to optimize the electrical delay of the drive signal of the PM${}_\textrm{in}$. The electrical signal applied to the PM${}_\textrm{in}$ is a square wave with a duty cycle of 50\% and a period of 100 ns. We scan the V${}_\textrm{pp}$ between 0 V and 3 V with a repetition of 0.2 Hz, corresponding to a period of 5 s. If both the early and late bin are at the high or low level simultaneously, scanning the drive voltage of PM${}_\textrm{in}$ will not change the relative phase $\alpha$ between the early and late bin. Consequently, there will be no interference at the output ports of AMZI and the visibility will be close to zero.

\begin{figure}[h]
	\includegraphics[width=0.8\linewidth]{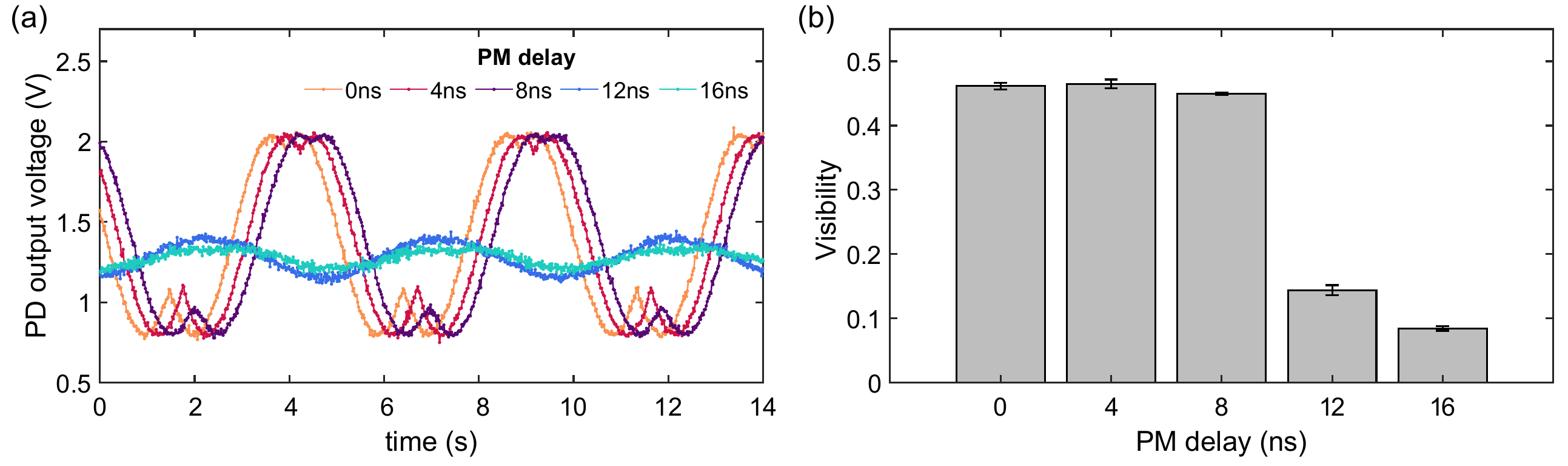}
      	\caption{\label{figS5}Optimize the delay of PM${}_\textrm{in}$. (a) The optical power intensity interference at the output of the analyzing AMZI when scanning the PM${}_\textrm{in}$'s voltage. (b) The visibility of the interference for different delays of PM${}_\textrm{in}$.}
	\end{figure}

    Conversely, if the early bin is at a high level and the late bin is at a low level (or vice versa), scanning the V${}_\textrm{pp}$ will result in an interference at two output ports of AMZI, and the visibility will be close to 50\% for an ideal case. In Fig. \ref{figS5}(b), we show the visibility for different delays of the PM${}_\textrm{in}$ and we chose 0 ns as the delay of the PM${}_\textrm{in}$ in the final experiment.

\section{Note S5: Optimize the temporal delay of IM${}_\textrm{in}$}
\begin{figure}[h]
	\includegraphics[width=0.8\linewidth]{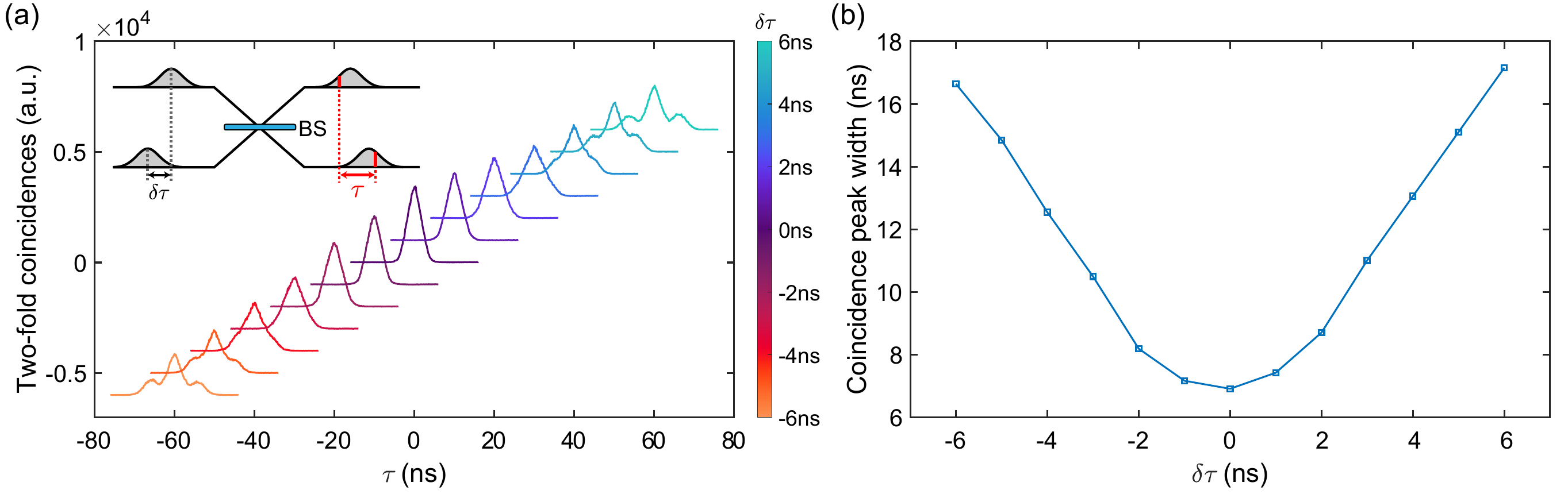}
      	\caption{\label{figS6}Optimize the delay of IM${}_\textrm{in}$. (a) The histogram of the two-fold coincidence at two outputs of the BS in the BSM. The color-bar indicates the difference in arrival time $\delta \tau$. Vertical and horizontal offset for clarity. (b) The width of the coincidence peak for different $\delta \tau$.}
	\end{figure}

    In the BSM, temporal indistinguishability is ensured by precisely controlling the arrival time of two photons interfered at the BS by two lithium niobate IMs driven by an AWG. As shown in the inset of Fig. \ref{figS6}(a), the input and idler photons are interfered at a BS with a relative delay of $\delta \tau$ and detected by two SNSPDs at two output ports of the BS. $\tau$ is the time difference between two photon detection events. If the relative difference in arrival time $\delta \tau$ is longer than the pulse duration of the photons, the histogram of two-fold coincidence counts will have three coincidence peaks. The central peak corresponds to the remaining two-fold coincidence counts caused by multiphoton events, and two side peaks are due to the accidental coincidence (A.C.) counts of the input and idler photons. By scanning the electrical delay of the IM${}_\textrm{in}$ with a step of 1 ns, we overlap the input photon and the idler photon in the time domain so that these three coincidence peaks gradually merge into one, as shown in Fig. \ref{figS6}(a). We fit the temporal widths of the center peak for various delays. As shown in Fig. \ref{figS6}(b), at $\delta \tau=0$, we obtain the narrowest peak, corresponding to the optimal delay of IM${}_\textrm{in}$.

\section{Note S6: Theoretical models for HOM interference and quantum beating}

\begin{figure}[h]
	    \includegraphics[width=0.8\linewidth]{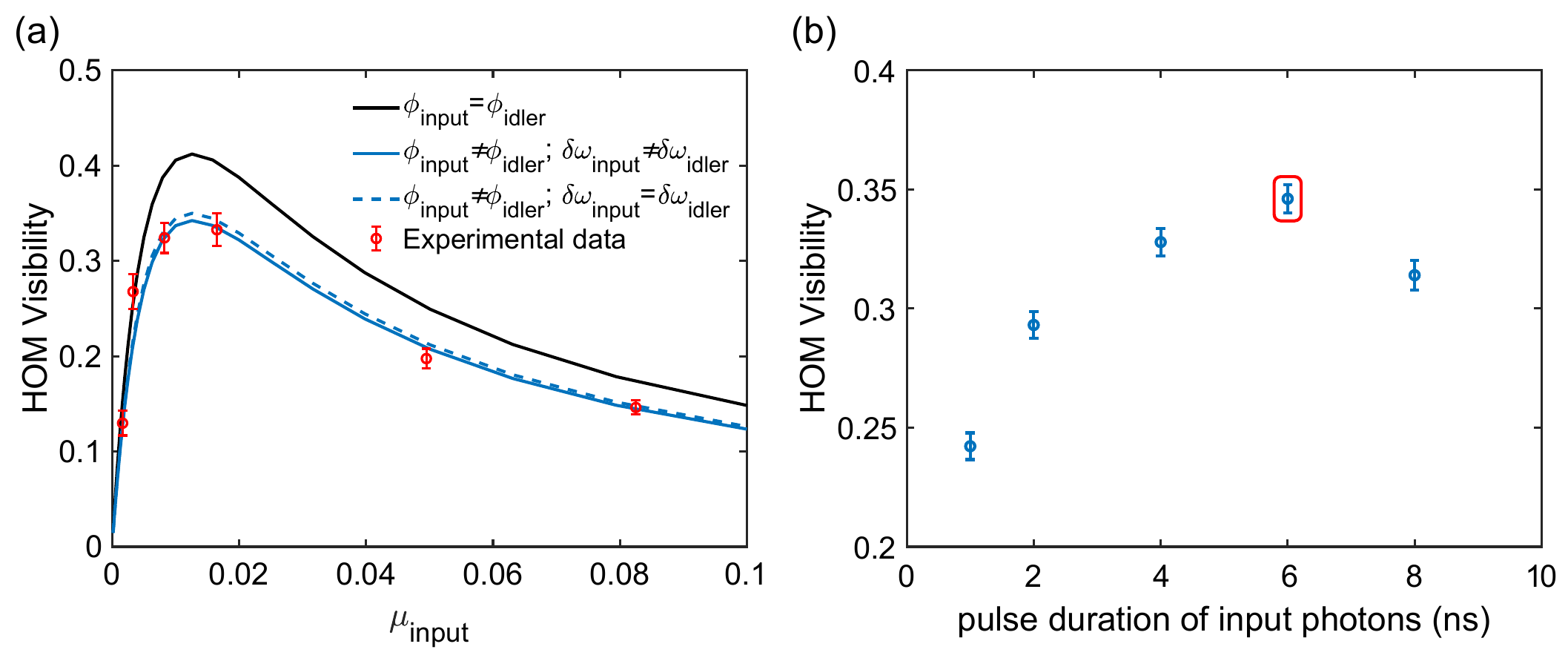}
      	\caption{\label{figS15} Experimental data and calculated result of HOM visibilities. (a) Calculated and measured HOM visibilities with varying mean photon numbers of idler photons; (b) Measured HOM visibilities with varying pulse durations of input photons.}
    \end{figure}

As we mentioned in the main text, one can obtain a maximum HOM visibility of 40$\%$ when interfering a coherent-state photon and a thermal-state photon. The calculation for ideal case, where both spectral lineshape and linewidth of the input and idler photons are equal, is shown in solid black curves in Fig. \ref{figS15}(a). However, the visibility measured in our experiment is 33.5$\%$, which indicates an indistinguishability of 83.7$\%$. We attribute the main reasons that limit the HOM visibility to the differences of both spectral lineshape and linewidth between the input and idler photons.

 First, we measure the HOM visibility with fixed mean photon numbers of idler photons while we vary that of input photon. The experimental result is shown as red circles in Fig. \ref{figS15}(a) and the calculation results with the same/different linewidth is shown in dashed/solid blue curves in Fig. \ref{figS15}(a). Note that due to the different spectral lineshape, the maximum of HOM visibility is about 35$\%$, instead of 40$\%$.
    
Second, we fix the linewidth of idler photons at 200 MHz and measure the HOM visibility with different pulse durations of input photons from 1 ns to 8 ns, corresponding to the linewidth from 150 MHz to 1200 MHz. As shown in Fig. \ref{figS15}(b), when the linewidth of input photons is close to idler, the visibility reaches a maximum of about 34.6$\%$, which agrees well with the calculation results mentioned above.
    
Combining the above analysis and calculations, we conclude that: the main factors that limit the HOM visibility are spectral lineshape and linewidth of input and idler photons. The difference in spectral lineshape decreases the HOM visibility from 40$\%$ to 35$\%$ and the difference in linewidth further reduces the visibility to 33.5$\%$.

To investigate the correlation between the HOM visibility and quantum bit error rate (QBER), we apply an analytical model introduced in \cite{valivarthi_quantum_2016}. In our experiment, the photon-number of input photon is Poissonian and thermal for idler. This matches with the same system that this model describes.

\begin{figure}[h]
	\includegraphics[width=0.8\linewidth]{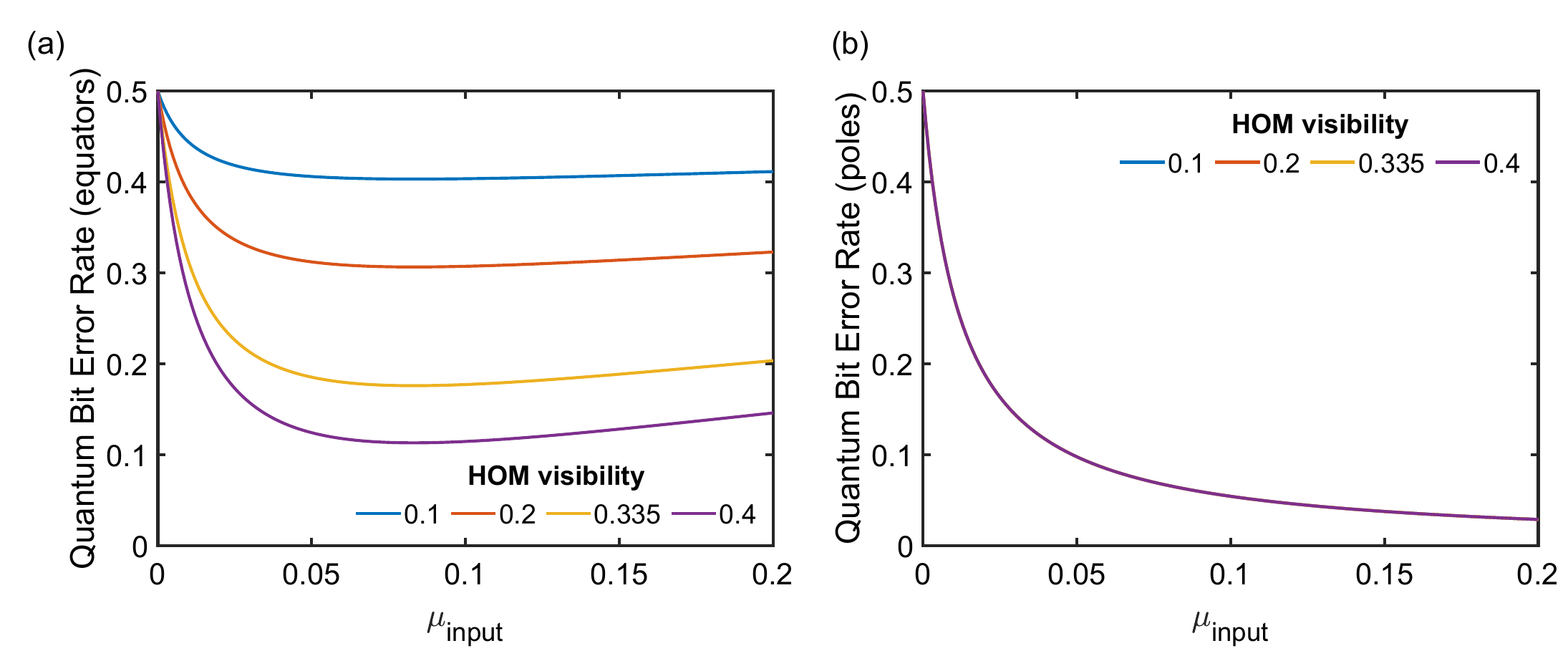}
      	\caption{\label{figS16} Calculated QBER with different HOM visibilities. (a) Equator qubits; (b) Poles qubits.}
    \end{figure}

First, the quantum bit error rate and the fidelity can be written as
        \begin{equation}
        \label{QBER}
            E = \frac{P_{\rm incorrect}}{P_{\rm correct} + P_{\rm incorrect}},
        \end{equation}
        \begin{equation}
        \label{fidelity}
            F = \frac{P_{\rm correct}}{P_{\rm correct} + P_{\rm incorrect}}
        \end{equation}
    where $P_{\rm incorrect}/P_{\rm correct}$ denotes the probability of incorrect/correct three-fold coincidence events between BSM and retrieved signal photons, respectively. For the equator qubit (Fig. \ref{figS16}(a)), the expected fidelity $F_{\rm equator}$ and QBER $E_{\rm equator}$ can be written as:
        \begin{equation}
        \label{QBER_eq}
            F_{\rm equator} = \frac{1}{2} + \frac{\varsigma}{2}\frac{P\left( {1,1,1} \right) + P\left( {1,1,2} \right)}{P\left( {1,1,1} \right) + P\left( {1,1,2} \right) + P\left( {2,0,1} \right) + P\left( {2,0,2} \right) + P\left( {0,2,1} \right) + P\left( {0,2,2} \right)},
        \end{equation}
        \begin{equation}
        \label{fidelity_eq}
        \begin{split}
            &E_{\rm equator} = 1 - F_{\rm equator} = \\&\frac{1}{2}\frac{\left( {1 - \varsigma} \right)\left( {P\left( {1,1,1} \right) + P(1,1,2)} \right) + \left( {P\left( {2,0,1} \right) + P\left( {2,0,2} \right) + P\left( {0,2,1} \right) + P(0,2,2)} \right)}{P\left( {1,1,1} \right) + P\left( {1,1,2} \right) + P\left( {2,0,1} \right) + P\left( {2,0,2} \right) + P\left( {0,2,1} \right) + P(0,2,2)}
        \end{split}
        \end{equation}
    where $P(l,m,n)$ are are the probabilities to detect a three-fold coincidence event with $l$ input photons and $m$ idler photons arriving at the BS, $n$ signal photons arriving at the detector. For example, $P(2,0,1)$ describes that one can detect a BSM event by both two photons coming from the input, while the idler photon is lost during the propagation and the signal photon arrives the detector successfully. $\varsigma$ represents the indistinguishability between input and idler photon, given by
        \begin{equation}
        \label{indisting}
            \varsigma = \frac{V({\rm experiment})}{V({\rm theory})}
        \end{equation}
        Similarly, we can also estimate the fidelity and QBER of pole qubits $F_{poles}$ and $E_{poles}$ (Fig. \ref{figS16}(b)) as:
        \begin{equation}
        \label{QBER_poles}
            F_{\rm poles} = \frac{P\left( {1,1,1} \right) + P\left( {1,1,2} \right) + \frac{1}{2}\left( {P\left( {0,2,2} \right) + P(0,2,1)} \right)}{P\left( {1,1,1} \right) + P\left( {1,1,2} \right) + P\left( {0,2,1} \right) + P(0,2,2)},
        \end{equation}
        \begin{equation}
        \label{fidelity_poles}
            E_{\rm poles} = \frac{\frac{1}{2}\left( {P\left( {0,2,2} \right) + P(0,2,1)} \right)}{P\left( {1,1,1} \right) + P\left( {1,1,2} \right) + P\left( {0,2,1} \right) + P(0,2,2)}
        \end{equation}
    Note that for $|{\rm e}\rangle$ or $|{\rm l}\rangle$, $P(2,0,1)$ and $P(2,0,2)$ will not result in a BSM event anymore and the fidelity and QBER will not be affected by the indistinguishability $\varsigma$.

    In frequency resolved Hong-Ou-Mandel interference, there is a time resolved quantum beating when the central frequencies of the input photon and the idler photon are not equal. In this note, we develop a theoretical model following the Ref. \cite{twocolor_Rarity_1990,QBeating_Ou_Mandel_1988,legero_time-resolved_2003,photonics_2023}. 
    In our experiment, the input state to be teleported is a weak coherent state prepared from attenuated laser, which can be represented as
    \begin{equation}
		\label{eqS14}
        |\alpha\rangle_\textrm{A}=\exp\Big(-\frac{|\alpha|^2}{2}\Big)\sum_{n}\frac{\alpha^n}{\sqrt{n!}}|n\rangle,
	\end{equation}
    in which $|\alpha|^2=\mu$ is the mean photon number. The $n$ photons probability distribution obeys Poissonian distribution, which is given by
    \begin{equation}
		\label{eqS15}
        P(\mu, n)=e^{-\mu}\frac{\mu^n}{n!}.
	\end{equation}
    Assuming the idler photon as a single-mode thermal state, the $n$ photons probability distribution obeys Bose-Einstein distribution:
    \begin{equation}
		\label{eqS16}
        P(\mu, n)=\frac{\mu^n}{(1+\mu)^{n+1}}.
	\end{equation}
    Therefore, when we interfere the input photons and idler photons at a BS, the probability of finding the photon number states $|m_c,n_t\rangle$ at the input ports of BS can be written as
    \begin{equation}
		\label{eqS17}
        P(m_\textrm{c}, n_\textrm{t}|\mu_\textrm{c},\mu_\textrm{t})=e^{-\mu_\textrm{c}}\frac{\mu_\textrm{c}^{m_\textrm{c}}}{m_\textrm{c}!}\frac{\mu_\textrm{t}^{n_\textrm{t}}}{(1+\mu_\textrm{t})^{n_\textrm{t}+1}},
	\end{equation}
    where the subscripts ``c'' for ``coherent'' and ``t'' for ``thermal'', and $\mu_\textrm{c}$ and $\mu_\textrm{t}$ is the mean photon number of coherent input photons and thermal idler photons, respectively. Thus, we can compute the probability of two-fold coincidence counts $P^{\mu_\textrm{c},\mu_\textrm{t}}$, given by
    \begin{equation}\label{eqS18}
    \begin{split}
        P^{\mu_\textrm{c},\mu_\textrm{t}}=&P(1_\textrm{c}, 1_\textrm{t}|\mu_\textrm{c},\mu_\textrm{t})P(1_\textrm{A},1_\textrm{B}|1_\textrm{t},1_\textrm{c})    \\
        &+P(2_\textrm{c}, 0_\textrm{t}|\mu_\textrm{c},\mu_\textrm{t})P(1_\textrm{A},1_\textrm{B}|2_\textrm{c},0_\textrm{t})    \\
        &+P(0_\textrm{c}, 2_\textrm{t}|\mu_\textrm{c},\mu_\textrm{t})P(1_\textrm{A},1_\textrm{B}|0_\textrm{c},2_\textrm{t})
    \end{split}
	\end{equation}
    where $P(1_\textrm{A},1_\textrm{B}|m_\textrm{c},n_\textrm{t})$ is the probability of two-fold coincidence counts between the output port A and B of the BS, conditioned on finding $m_c$ photons from the coherent input states and $n_t$ photons from the thermal idler states. 
    
    When we introduce a frequency detuning, $\Delta$, between input photons and idler photons, these two photons are partially distinguishable by their frequencies. Hence, the probability $P(1_\textrm{A},1_\textrm{B}|1_\textrm{c},1_\textrm{t})$ is not equal to zero anymore. We apply the analytical model in Ref.\cite{legero_time-resolved_2003} to consider this case. The spatial-temporal mode functions of input and idler photons can be written as
    \begin{equation}
    \begin{aligned}
		\label{eqS19}
        \zeta_\textrm{c}(t)=\sqrt[4]{\frac{2}{\pi}}\exp\bigg(-\Big(t-\frac{\delta \tau}{2}\Big)^2-i\Big(\omega-\frac{\Delta}{2}\Big)t\bigg) \\
        \zeta_\textrm{t}(t)=\sqrt[4]{\frac{2}{\pi}}\exp\bigg(-\Big(t+\frac{\delta \tau}{2}\Big)^2-i\Big(\omega+\frac{\Delta}{2}\Big)t\bigg)
        \end{aligned}
	\end{equation}
    where $\omega=\tfrac{\omega_1+\omega_2}{2}$. The coincidence probability at two output ports of BS is given by
    \begin{equation}
		\label{eqS20}
          P(1_\textrm{A},1_\textrm{B}|1_\textrm{c},1_\textrm{t})=\frac{\cosh(2\tau\delta\tau)-\cos(\tau\Delta)}{2\sqrt{\pi}}e^{-(\tau^2+\delta\tau^2)}.
	\end{equation}
    Here, $\tau$ is the detection-time delay between output port A and port B of the BS and $\delta\tau$ is the arrival-time delay between input and idler photon. Similarly, we can also calculate the probability $P(1_\textrm{A},1_\textrm{B}|2_\textrm{c},0_\textrm{t})$ and $P(1_\textrm{A},1_\textrm{B}|0_\textrm{c},2_\textrm{t})$:
    \begin{equation}
		\label{eqS21}
        P(1_\textrm{A},1_\textrm{B}|2_\textrm{c},0_\textrm{t})=P(1_\textrm{A},1_\textrm{B}|0_\textrm{c},2_\textrm{t})=\frac{1}{\sqrt{\pi}}e^{-\tau^2}.
	\end{equation}
    Substituting Eq. (\ref{eqS20}) and Eq. (\ref{eqS21}) into Eq. (\ref{eqS18}), the probability of two-fold coincidence counts finally becomes
    \begin{equation}\label{eqS22}
    \begin{split}
        P^{\mu_\textrm{c},\mu_\textrm{t}}=&\frac{1-\cos(\tau\Delta)}{2\sqrt{\pi}}e^{-\tau^2}\times e^{-\mu_\textrm{c}}\frac{\mu_\textrm{c}\mu_\textrm{t}}{(1+\mu_\textrm{t})^2} \\
        &+\frac{1}{\sqrt{\pi}}e^{-\tau^2}\times\bigg(e^{-\mu_\textrm{c}}\frac{\mu_\textrm{c}^2}{2}\frac{1}{1+\mu_\textrm{t}}+e^{-\mu_\textrm{c}}\frac{\mu_\textrm{t}^2}{(1+\mu_\textrm{t})^3}\bigg)
    \end{split}
	\end{equation}
     The theoretical fitting results based on Eq. (\ref{eqS22}) are shown in the Fig. 2 of the main text.

\section{Note S7: Details of the quantum memory and Experimental Time Sequence}
\begin{figure*}
  	\includegraphics[width=0.8\linewidth]{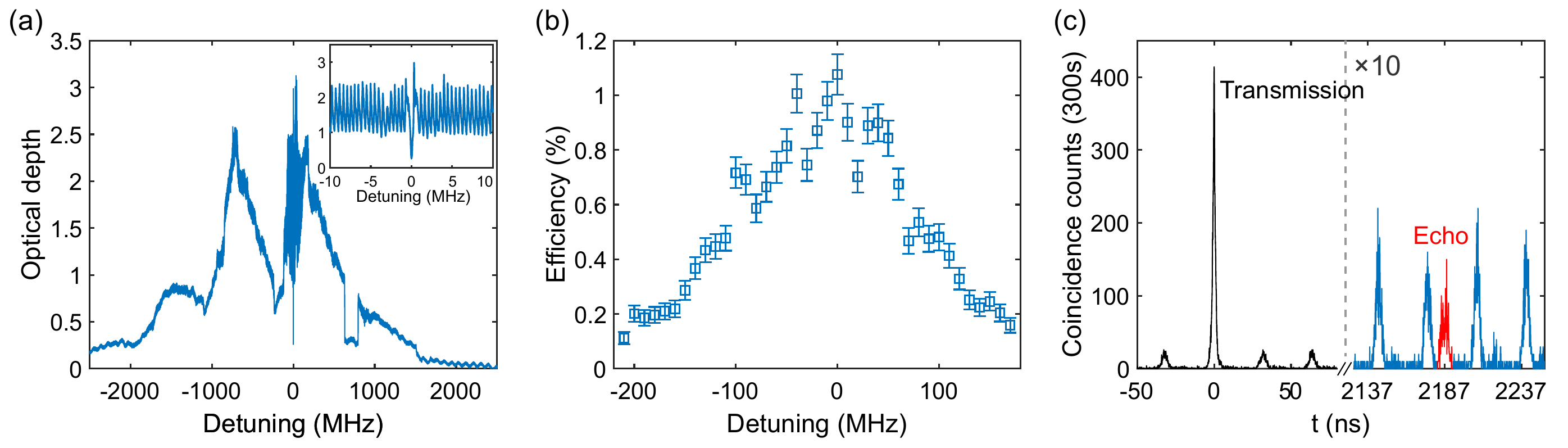}
  	\caption{\label{figS_AFC} An atomic-frequency-comb (AFC) quantum memory based on a ${\rm ^{167}Er^{3+}\!\!:\!\!Y_{2}SiO_{5}}$ crystal. (a) The full absorption spectrum of the $^4I_{15/2}(0)\ (m_{S}=-1/2)\leftrightarrow {}^4I_{13/2}(0)\ (m_{S}=-1/2)$ optical line with a 1.5 T magnetic field for 2187 ns storage time. Insets: the zoom-in views of observed optimal AFC. The center distortion is the result of the limited extinction ratio of the ${\rm IM_A}$. (b) Quantum memory efficiency with the detuning between the AFC central frequency and the signal photons. (c) The histogram of two-fold coincidence counts, vertically offset for clarity. The data beyond 2135 ns are magnified by a factor of 10.}
	\end{figure*}

\textbf{Quantum memory}: 
    The experimental scheme of the quantum memory is shown in Fig. 1(e) of the main text. We prepare an AFC structure between the $^4I_{15/2}(0)\ ( m_{S}=-1/2)\leftrightarrow {}^4I_{13/2}(0)\ (m_{S}=-1/2)$ transition at about $1536.17$ nm of the ${\rm ^{167}Er^{3+}}$. There are 8 hyperfine levels in the respective ground and excited  levels and thus the absorption profiles consist of 64 optical transitions with different hyperfine transitions, $\Delta m_{I}$. As the GHz-wide inhomogeneous broadening makes these hyperfine transitions unresolved, these transitions all participates in the pump process, and we implement a partial initialization to prepare the AFC as described below. First, we use the frequency-locked AFC laser and another carrier-suppression single-side-band modulator (CS-SSB$\rm {_{A}}$), shifting the frequency of the AFC-laser with a detuning of 6.58 GHz, and a frequency sweep window of 160 MHz to partially polarize the ground states population of the ${\rm ^{167}Er^{3+}}$ ensemble mainly by ${\Delta m_{I}=+1}$ passages (the right square-pit in Fig. \ref{figS_AFC}(a)), and enhance the absorption around the ${\Delta m_{I}=0}$ hyperfine transitions, where the AFC is created next. We use CS-SSB$\rm {_{A}}$ to further shift the AFC laser's center frequency at 7.28 GHz detuning and burn a nearly 200-MHz AFC structure in this enhanced region. All RF signals are generated from a signal generator and synchronized with the optical switch to control the pulse sequence. 
    
    With a polarization power of about 90 $\mu$W and an AFC peak power near 3 mW, we prepare an AFC with a period of $\textrm{1/2187 ns} = \textrm{0.457 MHz}$, as shown in Fig. \ref{figS_AFC}(a) and its inset. To align the frequency between the AFC and the signal photon, we fix the frequency of signal photons and scan the AFC center frequency while keeping the frequency difference between the polarization and the AFC-center frequency fixed at about 700 MHz. We optimize their frequency alignment by measuring the storage efficiency, which is obtained by measuring two-fold coincidence counts between the idler photon and the retrieved signal photon, as shown in Fig. \ref{figS_AFC}(b). The storage efficiency is calculated as  
    \begin{equation}
        \label{eq3}
        \eta=\frac{\rm echo}{\rm input}=\frac{\textrm{C.C.}({\rm echo})}{\textrm{C.C.}({\rm transmission})}\times\frac{\textrm{S.C.}({\rm transmission})}{\textrm{S.C.}({\rm input})}.
	\end{equation}
    where C.C.(echo) and C.C.(transmission) are the respective coincidence counts at $\tau$=$t_{\rm M}$ and $\tau=0$. S.C.(transmission) and S.C.(input) are the respective single-photon counts with/without the AFC's absorption. Fig. \ref{figS_AFC} shows the optimal coincidence histogram with an integration time of 300 s. The overall storage efficiency is about 1.1\%, which is only about half of the theoretical efficiency of 2.5$\%$ with a finesse of 3.2 for the observed optimal AFC (shown in Fig. \ref{figS_AFC}(b) and its inset). As we observed the fluctuation of the contrast of the AFC absorption spectrum on an oscilloscope, we attribute this discrepancy to the jitter of the energy level of ${\rm ^{167}Er^{3+}}$ due to the vibration of the crystal mount and the inhomogeneous magnetic field strength, which need further verification.
    \begin{figure}[h]
	\includegraphics[width=0.8\linewidth]{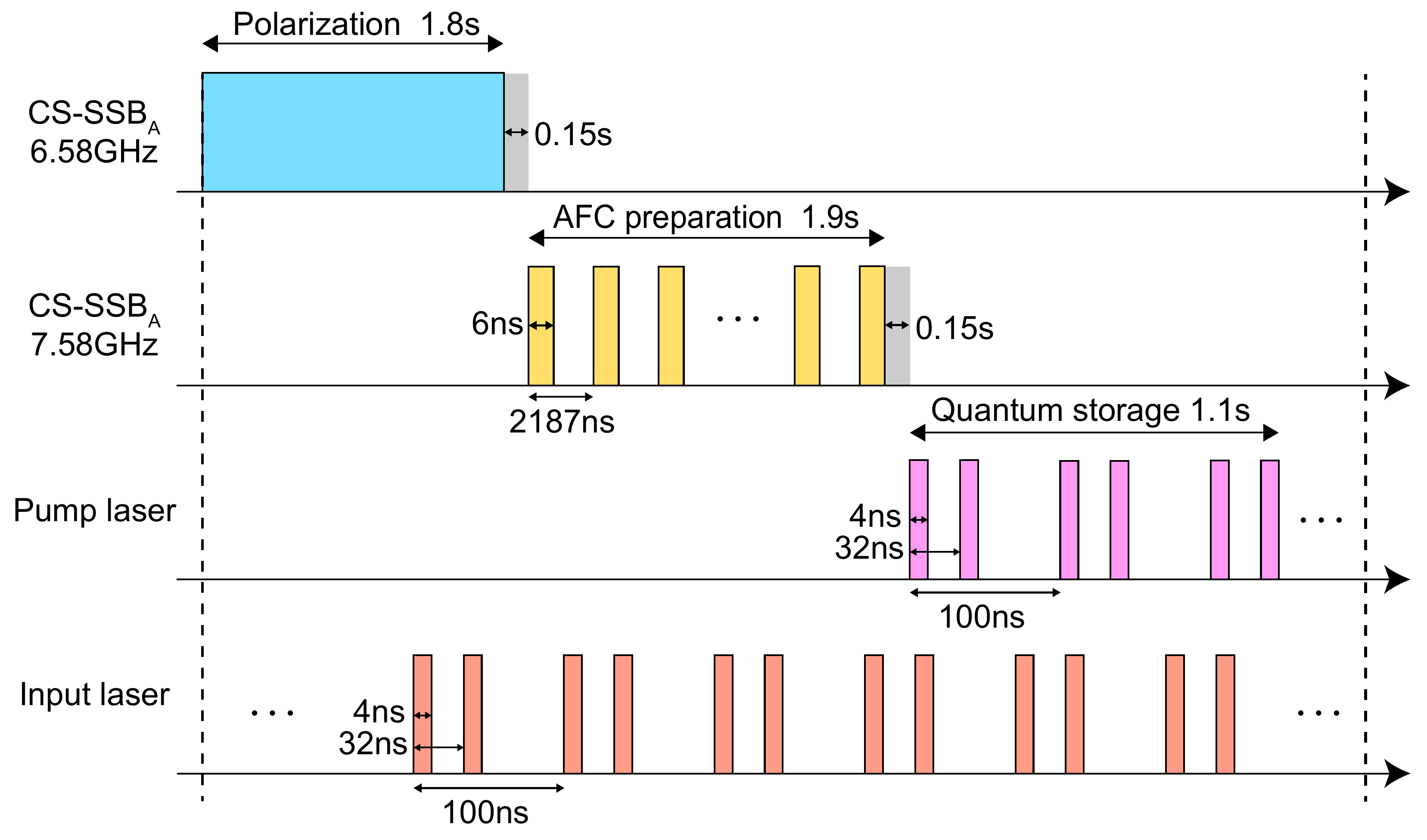}
      	\caption{\label{figS7}Experimental time sequence.}
	\end{figure}
 
    \textbf{Experimental time sequence}:
    The time sequence of the experiment is shown in Fig. \ref{figS7}. Firstly, to partially polarize the ions and increase the optical depth, the frequency of the AFC laser is shifted by CS-SSB${}_\textrm{A}$ with a 6.58-GHz detuning and a sweep range of 160 MHz for 1.8 s. Secondly, CS-SSB${}_\textrm{A}$ shifts the frequency of AFC laser with a 7.28-GHz detuning, and the AFC laser is intensity modulated into pulses with a duration of 6 ns and a period of 2187 ns, equal to the storage time. The whole AFC preparation process continues for 1.9 s. Finally, after 0.15-s waiting time, the signal photons are sent to the prepared quantum memory, and then retrieved for qubit analysis. Throughout the whole experiment, an arbitrary waveform generator is used to provide modulation signals to the AFC preparation, EPR source pump and IM${}_\textrm{in}$ for input state preparation respectively. A signal generator is employed to drive the CS-SSB${}_\textrm{A}$. The optical switch controls the optical path into the refrigerator during the AFC preparation and the storage and analysis process. See Fig. 1 of the main text for details.

\section{Note S8: Impacts of spin-wave AFC on the entanglement distribution rates}
\begin{figure*}
  	\includegraphics[width=0.8\linewidth]{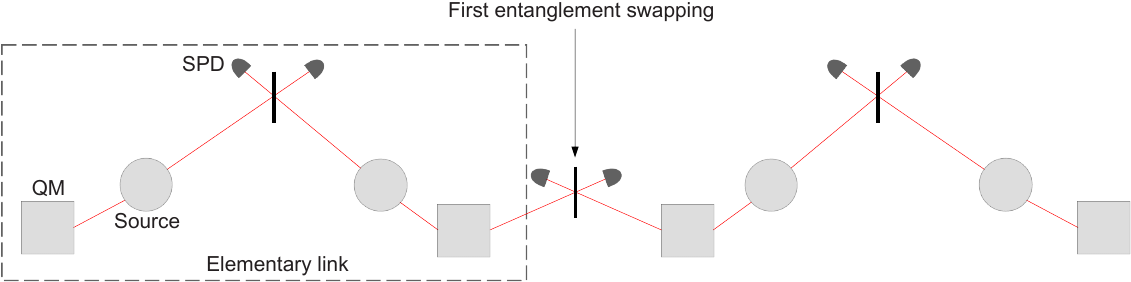}
  	\caption{\label{figS_qr} A quantum repeater consists of two elementary links. Entanglement distribution between two elementary links is achieved by entanglement swapping.}
	\end{figure*}

    As a further step based on the two-level AFC, the spin-wave storage version of AFC has advantages in the distribution of entanglement for its longer storage time and on-demand retrieval property. For the entanglement distribution inside the elementary link\cite{RMP_2011}, which is composed of two pairs of entanglement source and multimode quantum memory located at two nodes with a distance of $L_0$ (Fig. \ref{figS_qr}). The heralding rate $R_{\rm heralding}$ is given by
    \begin{equation}
    \frac{c}{L_0}\times\big[1-\big(1-R_{\rm idler}\times\Delta t\times\eta_{\rm IL}^{\rm idler}\big)^M\big]\times M_{\rm time}\times\eta_{\rm duty}
    \end{equation}
    for single photon detection protocol\cite{lago-rivera_telecom-heralded_2021,hänni2025}, and
    \begin{equation}
    \frac{c}{L_0}\times\big[1-\big(1-\frac{1}{2}\times(R_{\rm idler}\times\Delta t\times\eta_{\rm IL}^{\rm idler})^2\big)^M\big]\times M_{\rm time}\times\eta_{\rm duty}
    \end{equation}
    for two photon detection protocol\cite{Spectral_Mulplex_PRL_2014}, where $R_{\rm idler}$ is the idler rate from the source, $\Delta t$ is the detection window, $M$ is the multimode number other than temporal multiplex with mode number of $M_{\rm time}$, $\eta_{\rm IL}^{\rm idler}$ is the insertion losses including detection efficiency for idler, and $\eta_{\rm duty}$ is the duty cycle of the memory as the period of quantum memory including preparation is usually much longer than the trial period of $L_0/c$. The respective entanglement distribution rates between the two memories in an elementary link for the two protocols\cite{Spectral_Mulplex_PRL_2014,lago-rivera_telecom-heralded_2021,hänni2025} is:
    \begin{equation}
    \frac{c}{L_0}\times\big[1-\big(1-R_{\rm idler}\times\Delta t\times\eta_{\rm IL}^{\rm idler}\big)^M\big]\times \eta_{\rm IL}^{\rm signal}\times \eta_{\rm storage}^{\rm signal}\times M_{\rm time}\times\eta_{\rm duty}
    \end{equation}
    and
    \begin{equation}
    \frac{c}{L_0}\times\big[1-\big(1-\frac{1}{2}\times(R_{\rm idler}\times\Delta t\times\eta_{\rm IL}^{\rm idler})^2\big)^M\big]\times(\eta_{\rm IL}^{\rm signal})^2\times (\eta_{\rm storage}^{\rm signal})^2\times M_{\rm time}\times\eta_{\rm duty},
    \end{equation}
    where $\eta_{\rm IL}^{\rm signal}$ is the insertion loss including detection efficiency for signal, $\eta_{\rm storage}^{\rm signal}$ is the quantum storage efficiency.
    
    Adding spin-wave storage in this elementary link will increase memory time and hence increase the distance between two memories. Besides, the long-time storage realized on the hyperfine spin levels and the on-demand property are crucial for achieving high entanglement distribution rate in the second level of quantum network (i.e., first entanglement swapping in \cite{RMP_2011}, shown in Fig. \ref{figS_qr}).
    
    For example, considering a basic quantum repeater consists of two elementary links\cite{PRL_Simon_2007,RMP_2011,Spectral_Mulplex_PRL_2014,lago-rivera_telecom-heralded_2021,hänni2025}, the entanglement distribution rate $R_{\rm EPR}$ is given by
    \begin{equation}
    \frac{c}{L_0}\times\big[1-\big(1-R_{\rm idler}\times\Delta t\times\eta_{\rm IL}^{\rm idler}\big)^M\big]^a\times \eta_{\rm IL}^{\rm signal}\times \eta_{\rm storage}^{\rm signal}\times \frac{1}{2}\times M_{\rm time}\times\eta_{\rm duty}
    \end{equation}
    and 
    \begin{equation}
    \frac{c}{L_0}\times\big[1-\big(1-\frac{1}{2}(R_{\rm idler}\times\Delta t\times\eta_{\rm IL}^{\rm idler})^2\big)^M\big]^a\times(\eta_{\rm IL}^{\rm signal})^2\times (\eta_{\rm storage}^{\rm signal})^2\times \frac{1}{2}\times M_{\rm time}\times\eta_{\rm duty}.
    \end{equation} 
    The number $a=2$ for quantum memory without spin-wave storage, and $a=1$ with the on-demand spin-wave storage memory as one link that has been heralded in one temporal mode can wait until the other link is successfully heralded in another temporal mode. Therefore, the spin-wave storage version of AFC enables higher entanglement distribution rate in quantum repeater protocol than the sole optical AFC storage.
 
\section{Note S9: Quantum state tomography}
\begin{figure}
	\includegraphics[width=0.8\linewidth]{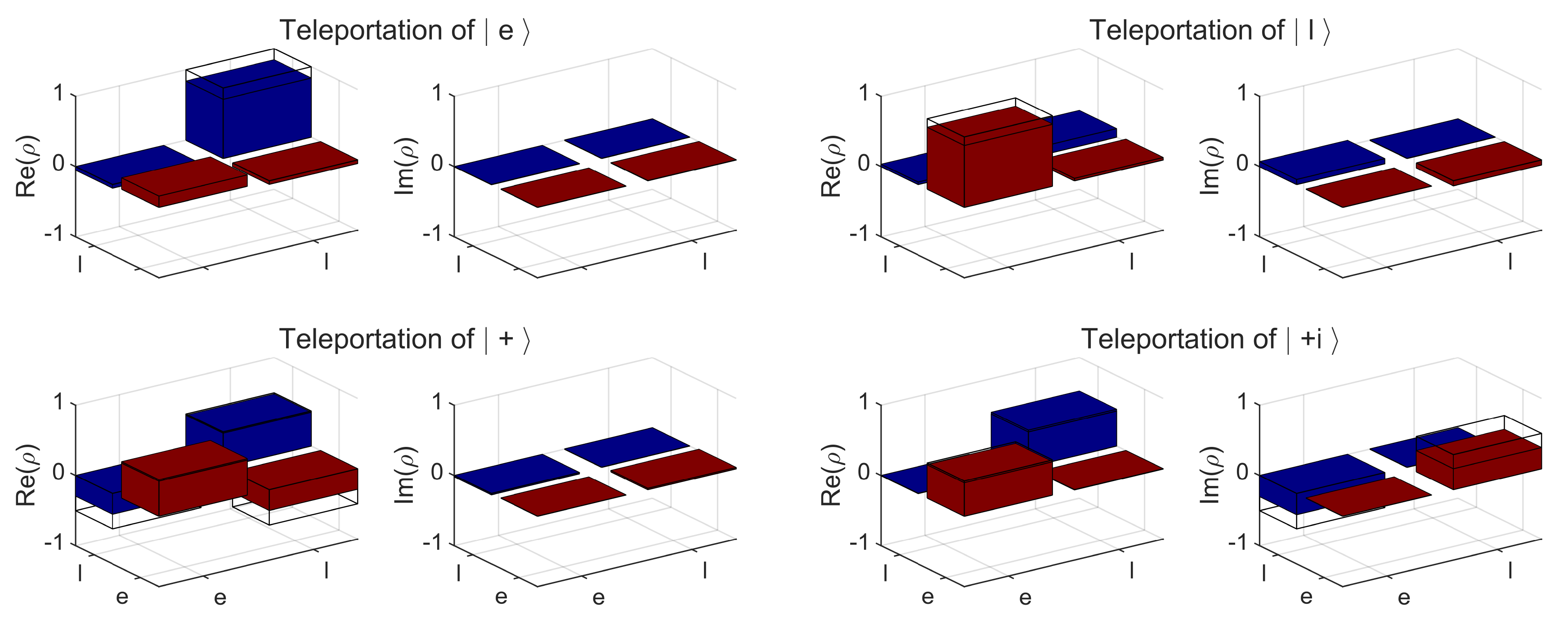}
      	\caption{\label{figS8}Density matrices reconstructed with quantum state tomography. Quantum state tomography results of the four quantum states with the BSM outcome of $|\Psi^{-}\rangle$. The bar graphs show the reconstructed real and imaginary parts of the density matrices for the four states teleported from telecom photons to erbium-ion ensembles. The wire grid indicates the expected values for the ideal cases. The data shown comprise a total of 790 three-fold coincidence counts in about 12 h. The uncertainties in state fidelities extracted from these density matrices are calculated using a Monte Carlo routine using Poissonian errors.}
	\end{figure}

    We perform quantum state tomography (QST)\cite{Daniel_PRA_2001} to reconstruct the density matrix of the stored and retrieved signal photons from quantum memory after teleportation. The density matrix can be written as
    \begin{equation}
		\label{eqS23}
        \hat{\rho}=\frac{1}{2}\sum_{i=0}^{3}S_{i}\sigma_{i},
	\end{equation}
    where $\sigma_{i}$ is the Pauli matrix and $S_{i} (i=0,1,2,3)$ are the four Stokes parameters which can be calculated by projection measurements on $X$, $Y$, $Z$ basis.
    \begin{equation}
    \begin{aligned}
		\label{eqS24}
        S_0&=p_\textrm{e}+p_\textrm{l} \\
        S_1&=p_\textrm{+}-p_\textrm{-} \\
        S_2&=p_{\textrm{+i}}-p_{-\textrm{i}} \\
        S_3&=p_\textrm{e}-p_\textrm{l}
    \end{aligned}
	\end{equation}
    Here $p_\textrm{e}$ and $p_\textrm{l}$ are the projection measurement results on $Z$-basis, $p_\textrm{+}$ and $p_\textrm{-}$ are the results for $X$-basis and $p_{\textrm{+i}}$ and $p_{\textrm{-i}}$ are the results for $Y$-basis, respectively. With these results, we can obtain the reconstructed density matrix of the stored and retrieved photons after teleportation, as shown in Fig. \ref{figS8}, and further calculate the state fidelities $F=\langle\psi_\textrm{B}|\hat{\rho}_\textrm{out}|\psi_\textrm{B}\rangle$ with the ideal states $|\psi_\textrm{B}\rangle$.
 
\section{Note S10: Optimization of the mean photon numbers}
\begin{figure}[h]
	\includegraphics[width=0.6\linewidth]{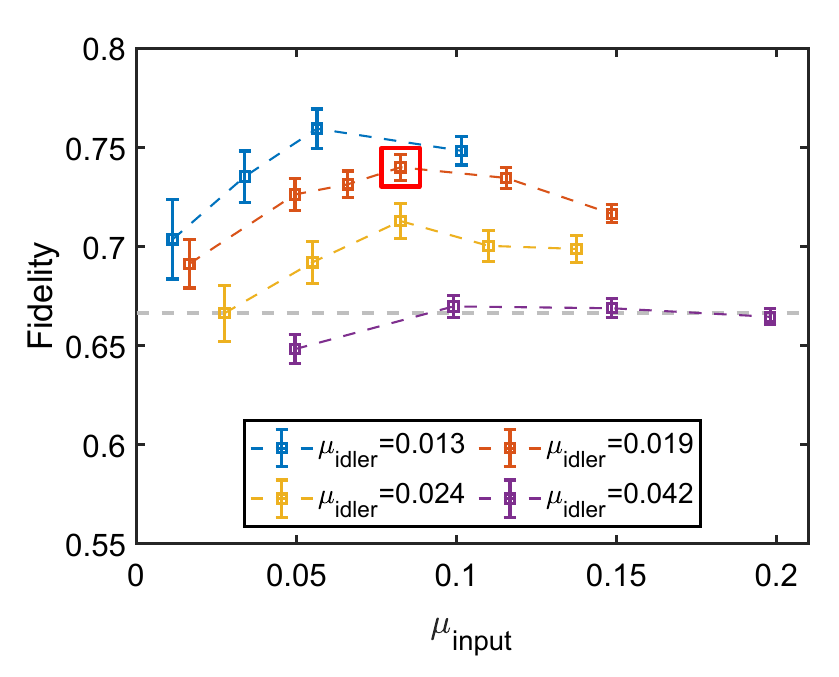}
      	\caption{\label{figS9}Fidelity as a function of $\mu_\textrm{input}$, for different $\mu_\textrm{idler}$. The highlighted point corresponds to the optimized experimental conditions: $\mu_\textrm{input}=0.0825$ and $\mu_\textrm{idler}=0.019$. The error bars correspond to $\pm1\sigma$.}
	\end{figure}

    In the experiment, the mean photon number of input photons ($\mu_\textrm{input}$) and the mean photon number of idler photons ($\mu_\textrm{idler}$) are determined as a trade-off between counting rates and teleportation fidelity. Due to the nature of SFWM, on one hand, higher pump power leads to higher $\mu_\textrm{idler}$ and the enhancement of coincidence counts; on the other hand, multi-photon events from SFWM decrease the teleportation fidelity. Here, the fidelities of input $|+\rangle$ states for different $\mu_\textrm{idler}$ as a function of $\mu_\textrm{input}$ are illustrated in Fig. \ref{figS9}. In the final teleportation experiment, we set the on-chip pump power of 1.48 mW, corresponding to a $\mu_\textrm{idler}$ of 0.019. And we set the mean photon number of input photons $\mu_\textrm{input}$ to 0.0825 for maximum fidelity with reasonable count rate.
    
\section{Note S11: Upper bound of fidelity using a classical strategy}
    In the main text, we show that the measured fidelity exceeds the maximum classical bound of 2/3, which is applied to input states encoded in a single photon. In our experiment, we prepared the input states from an attenuated laser with Poissonian statistics. In order to demonstrate the quantum nature of our teleportation system, we calculate the theoretical upper bound of fidelity using a classical strategy following Ref. \cite{PolStorage_PRL_2012}. 

    In a purely classical strategy, Alice performs a measurement on the input photon and Bob could prepare a completely new qubit based on Alice's measurement results. The low efficiency of the quantum channel and the multi-photon components of the weak coherent state allow Alice to obtain more information about the input state leading to a higher classical limit. It has been shown that for an $N$-photon qubit encoded in a weak coherent state, the best fidelity of the classical strategy is given by\cite{MassarPRL1995,specht_single-atom_2011}
    \begin{equation}
		\label{eqS25}
        F_\textrm{class}=\sum^{\infty}_{N\geq 1}\frac{N+1}{N+2}\frac{P(\mu,N)}{1-P(\mu,0)},
	\end{equation}
    where $P(\mu,N)=\dfrac{e^{-\mu}\mu^N}{N!}$ is the probability of $N$-photon in a coherent state with mean photon number of $\mu$. In our experiment, the mean photon number of the input state is 0.0825. Eq. (\ref{eqS25}) is only valid for an ideal quantum process with unity efficiency. In our work, this efficiency of quantum process is defined by the heralding probability of a retrieved signal photon conditioned on a successful BSM, which is
    \begin{equation}
		\label{eqS30}
        \eta=\frac{\textrm{C.C.(BSM, retrieved signal)}}{\textrm{C.C.(BSM)}}=5.50\times10^{-5}.
	\end{equation}
    \begin{figure}[h]
	\includegraphics[width=0.8\linewidth]{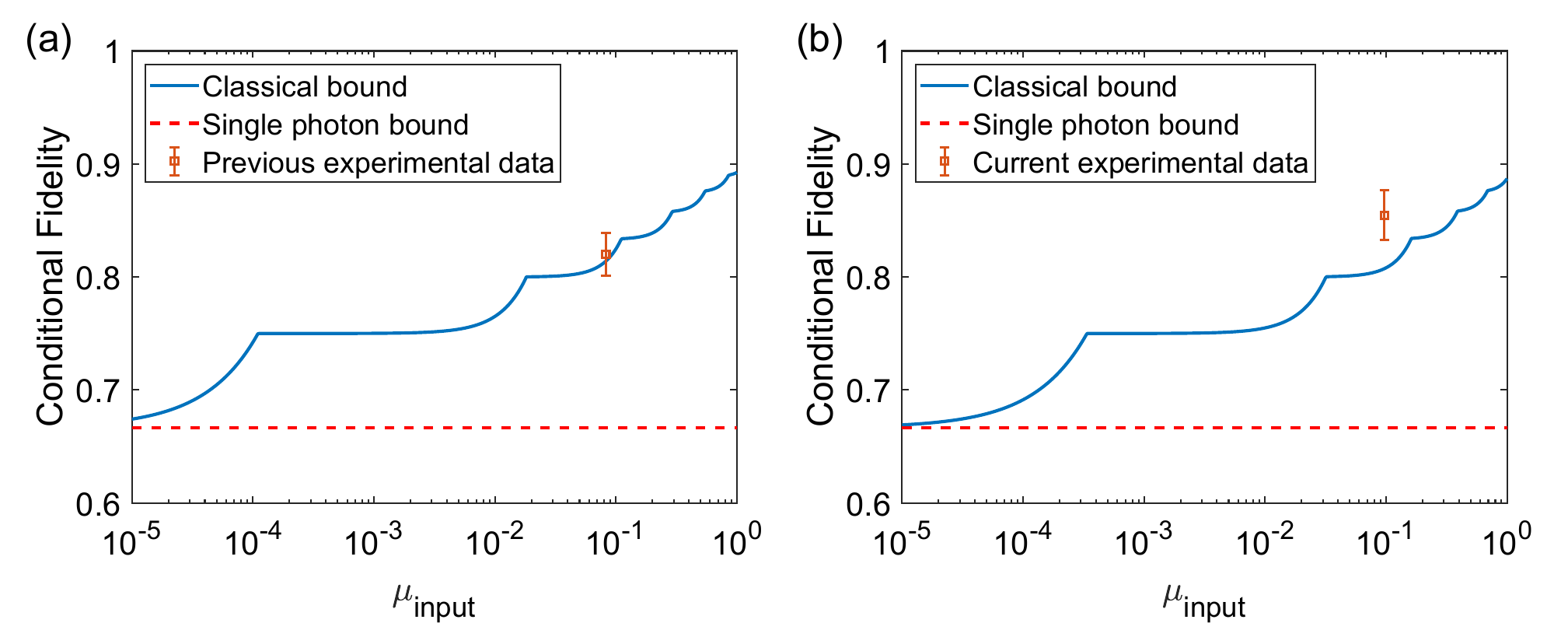}
      	\caption{\label{figS10}Classical limit of fidelity as a function of $\mu_\textrm{input}$, (a) Old photon-pair source; (b) New photon-pair source.}
	\end{figure}
 
    Note that the low efficiency is mainly due to the finite storage efficiency $\eta_\textrm{storage}=1.1\%$, the optical transmission loss $\eta_\textrm{t}=0.0956$ and the heralding efficiency of DMZI-R $\eta_\textrm{h}^\textrm{s}=\textrm{C.C.}(\textrm{idler, signal})/\textrm{S.C.}(\textrm{idler})=0.058$. In this case, the eventual efficiency can be written as
    \begin{equation}
		\label{eqS31}
        \eta=\eta_\textrm{storage}\eta_\textrm{t}\eta_\textrm{h}^\textrm{s}=6.09\times10^{-5},
	\end{equation}
    which is close to the experimental results. The classical maximum bound is plotted in Fig. \ref{figS10}(a). For $\mu_\textrm{input}$ = 0.0825, the maximum fidelity of classical strategy is 0.812 and the point corresponds to the experimental data, $F = 0.818\pm0.019 $. 
    
\begin{figure}[h]
	\includegraphics[width=0.6\linewidth]{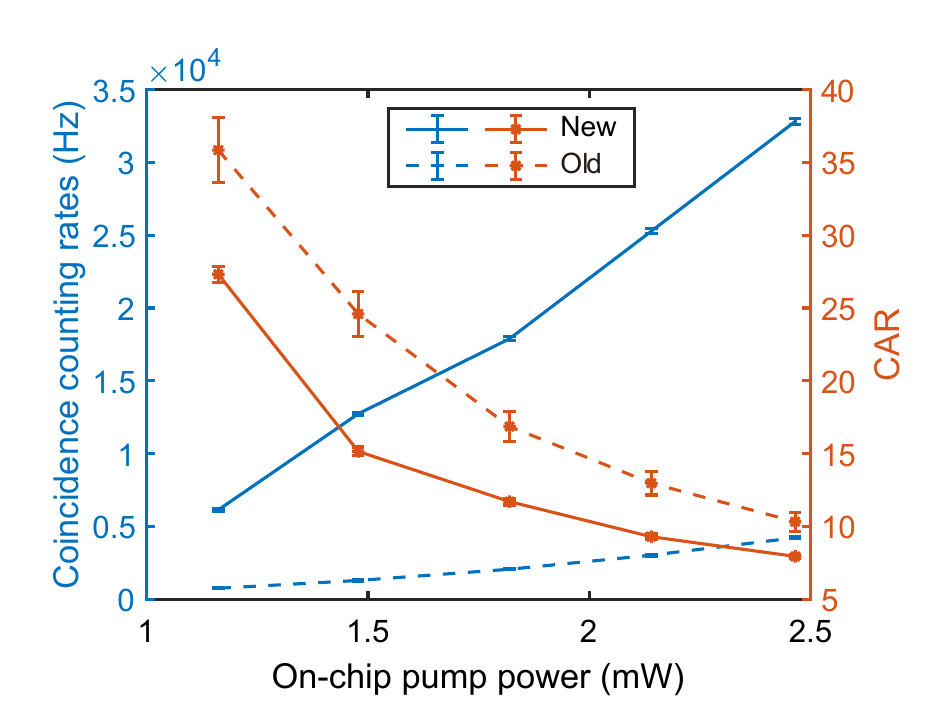}
      	\caption{\label{figS13} Comparison of the coincidence counting rate and CAR of the old and new photon-pair source.}
	\end{figure}

In order to further increase the teleportation fidelity, we improve the performance of our integrated photon-pair source and obtain 5.5-fold enhancement for coincidence counting rate under the same CAR. Comparison of the coincidence counting rate and CAR of the old and new photon-pair source is shown in Fig. \ref{figS13}. We now use a lower pump power of about 0.82 mW to suppress the multi-photon events. With the measured visibilities for two output ports of the analyzing AMZI, we calculate the fidelity for four different input states as 
    \begin{equation}
		\label{fid_vis}
        F=\frac{1+V}{2}
	\end{equation}
    and the results are listed in Table. {\ref{table2}}. The average fidelity of an arbitrary state is  $\bar{F}=\frac{1}{6}(F_{\rm e}+F_{\rm l}+2F_{\rm +}+2F_{\rm +i})$. Now we achieve an average fidelity of about $0.854\pm0.022 $.
 
 Moreover, the higher chip-to-fiber coupling efficiency of the new photon-pair source leads to a lower transmission loss for signal photons. The heralding probability of detecting a retrieved signal photons after the QM per BSM now is 
    \begin{equation}
		\label{eta_new}
        \eta_\textrm{new}=\frac{\textrm{C.C.(BSM, retrieved signal)}}{\textrm{C.C.(BSM)}}=1.68\times10^{-4},
	\end{equation}
    which is 3-times higher than that, $5.50\times10^{-5}$ of the old source. With a higher heralding probability, we could further reduce the upper bound of classical fidelity, which is now about 0.807. Our new results exceed the classical limit by more than 2 standard deviations. The calculation and experimental results are shown in Fig. \ref{figS10}(b).

    \begin{table}[h]
	\begin{ruledtabular}
	\begin{tabular}{cccc}
    	Input state  &  Visibility  &   Fidelity   &   Average fidelity  \\
    	\hline
    	$|{\rm e}\rangle$   &  0.776$\pm$0.021  &  0.888$\pm$0.011  &   \multirow{4}{*}{0.854±0.022}   \\
        $|{\rm l}\rangle$	  &  0.725$\pm$0.023  &  0.862$\pm$0.012  &      \\
    	$|{\rm +}\rangle$   &  0.690$\pm$0.056  &  0.845$\pm$0.028  &      \\
        $|{\rm +i}\rangle$  &  0.687$\pm$0.054 &  0.844$\pm$0.027  &      \\
  	\end{tabular}
   
	\end{ruledtabular}
    \caption{\label{table2}Experimental quantum state fidelities obtained with the new photon-pair source.}
	\end{table}
 
\section{Note S12: Calculate the fidelity of pure single photon with decoy state method}
    Following the Ref. \cite{ma_practical_2005,valivarthi_quantum_2016}, we estimate the fidelity of single-photon component in our experiment using decoy state method (DSM). DSM is usually used in quantum key distribution (QKD) to overcome the photon-number splitting attacks with decoy states\cite{ma_practical_2005,Decoy_PRL_2005}, which are encoded on coherent states with mean photon number $\mu_\textrm{d}$. In the teleportation, the DSM allows us to calculate the lower bound on the fidelity of the single-photon component in our teleportation system.

    First, we prepare two decoy states $|+\rangle_\textrm{d1}$ and $|+\rangle_\textrm{d2}$ and a signal state $|+\rangle_\textrm{s}$ with mean photon number of $\nu_1$, $\nu_2$ and $\mu_\textrm{s}$. The gain and quantum bit error rate (QBER) are given by\cite{ma_practical_2005}
    \begin{equation}\label{eqS32}
        Q^{(\mu)}=\sum_{n=0}^{\infty}\frac{Y^{(n)}\mu^{n}e^{-\mu}}{n!} \\
    \end{equation}  
    \begin{equation}\label{eqS33}
    E^{(\mu)}Q^{(\mu)}=\sum_{n=0}^{\infty}\frac{E^{(n)}Y^{(n)}\mu^{n}e^{-\mu}}{n!}	
    \end{equation}

    Here, $E^{(n)}$ and $Y^{(n)}$ is the error rate and the yield for a $n$-photon input, respectively. The lower bound of $Y^{(1)}$ and the upper bound of $E^{(1)}$ is given by
    \begin{equation}
		\label{eqS34}
        \begin{split}
        Y^{(1)}\geq Y^{(1)}_\textrm{Lower}&=\frac{\mu_\textrm{s}}{\mu_\textrm{s}\nu_1-\mu_\textrm{s}\nu_2-\nu_1^2+\nu_2^2}\\
        &\times\bigg(Q^{(\nu_1)}e^{\nu_1}-Q^{(\nu_2)}e^{\nu_2}-\frac{\nu_1^2-\nu_2^2}{\mu_\textrm{s}^2}\Big(Q^{(\mu_\textrm{s})}e^{\mu_\textrm{s}}-Y^{(0)}_\textrm{Lower}\Big)\bigg)\\
        \end{split}
	\end{equation}
    \begin{equation}
		\label{eqS35}
        \begin{split}
        E^{(1)}\leq E^{1}_\textrm{Upper}=\frac{E^{(\nu_1)}Q^{(\nu_1)}e^{\nu_1}-E^{(\nu_2)}Q^{(\nu_2)}e^{\nu_2}}{(\nu_1-\nu_2)Y^{(0)}_\textrm{Lower}},
        \end{split}
	\end{equation}
    where $Y^{(0)}_\textrm{Lower}$ is given by
    \begin{equation}
		\label{eqS36}
        Y^{0}\geq Y^{0}_\textrm{Lower}=\max\bigg{\{}\frac{\nu_1 Q^{(\nu_2)}e^{\nu_2}-\nu_2 Q^{(\nu_1)}e^{\nu_1}}{\nu_1-\nu_2},0\bigg{\}}
	\end{equation}
    According to Eq. (\ref{eqS35}), we can calculate a lower bound $F_\textrm{Lower}^1$ on the teleportation fidelity if we use a true single-photon to encode the input qubits
    \begin{equation}
		\label{eqS37}
        F^{(1)}=1-E^{(1)}\geq 1-E^{1}_\textrm{Upper}\equiv F_\textrm{Lower}^1.
	\end{equation}
    With the measurement results shown in Table. S1, the lower bound of $F^{(1)}$ for equatorial states $|+\rangle$ is $0.818\pm0.0125$, which is significantly exceed the classical limit of $2/3$ by about 10 standard deviations. The result demonstrates the quantum nature of our teleportation system.
    
    \begin{table}[t]
	\begin{ruledtabular}
	\begin{tabular}{cccc}
    	Input state  &  Mean photon number  & Gains  &  $F^{\mu}$  \\
    	\hline
    	Signal state  &  $\mu_\textrm{s}=0.0825$  &  $1.1\times10^{-6}$  & $0.740\pm0.0064$ \\
        decoy state 1  &  $\nu_1=0.0495$  &  $7.8\times10^{-7}$  &  $0.726\pm0.0079 $   \\
    	decoy state 2  &  $\nu_2=0.0165$  &  $3.2\times10^{-7}$  &  $0.691\pm0.0124$  
  	\end{tabular}
	\end{ruledtabular}\caption{\label{table1}}
	\end{table}

\bibliography{Teleportbib.bib}
\end{document}